\documentclass{article}
\usepackage{PRIMEarxiv}

\usepackage[utf8]{inputenc}
\usepackage[T1]{fontenc}
\usepackage{amsmath}
\usepackage{hyperref}
\usepackage{cleveref}
\usepackage{url}
\usepackage{booktabs}
\usepackage{amsfonts}
\usepackage{nicefrac}
\usepackage{microtype}
\usepackage{graphicx}
\usepackage{soul}
\usepackage{xcolor}
\sethlcolor{yellow}
\usepackage{subcaption}
\usepackage{makecell}
\usepackage[square,authoryear]{natbib}
\usepackage{enumitem} 
\usepackage{tcolorbox}

\newenvironment{significance}{\section*{Significance Statement}}{}

\hypersetup{
    colorlinks=true,
    citecolor=blue,
    linkcolor=blue,
    urlcolor=blue
}

\title{Conversational AI increases political knowledge as effectively as self-directed internet search}

\author{
\textbf{Lennart Luettgau\textsuperscript{1*\dag}} \quad
\textbf{Hannah Rose Kirk\textsuperscript{1*\dag}} \quad
\textbf{Kobi Hackenburg\textsuperscript{1*\dag}} \\
\textbf{Jessica Bergs\textsuperscript{1}} \quad
\textbf{Henry Davidson\textsuperscript{1}} \quad
\textbf{Henry Ogden\textsuperscript{2}} \quad
\textbf{Divya Siddarth\textsuperscript{3}} \quad
\textbf{Saffron Huang\textsuperscript{4}} \\
\textbf{Christopher Summerfield\textsuperscript{1\dag}} \\
\\
\textsuperscript{1}UK AI Security Institute, London, UK\\
\textsuperscript{2}AI Policy Directorate, London, UK\\
\textsuperscript{3}Collective Intelligence Project, San Francisco, CA, USA\\
\textsuperscript{4}Anthropic, San Francisco, CA, USA
}

\begin{document}
\maketitle
\renewcommand{\thefootnote}{\fnsymbol{footnote}}
\footnotetext[1]{These authors contributed equally to this work.}
\footnotetext[2]{Correspondence: lennart.luettgau@dsit.gov.uk hannah.kirk@dsit.gov.uk kobi.hackenburg@dsit.gov.uk christopher.summerfield@dsit.gov.uk}

\begin{abstract}
Conversational AI systems are increasingly being used in place of traditional search engines to help users complete information-seeking tasks. This has raised concerns in the political domain, where biased or hallucinated outputs could misinform voters or distort public opinion. However, in spite of these concerns, the extent to which conversational AI is used for political information-seeking, as well the potential impact of this use on users' political knowledge, remains uncertain. Here, we address these questions: First, in a representative national survey of the UK public (\textit{N} = 2,499), we find that in the week before the 2024 election as many as 32\% of chatbot users -- and 13\% of eligible UK voters -- have used conversational AI to seek political information relevant to their electoral choice. Second, in a series of randomised controlled trials (\textit{N} = 2,858 total) we find that across issues, models, and prompting strategies,
task-directed conversations with AI to research specific
political topics
%} 
increase political knowledge (increase belief in true information and decrease belief in misinformation) to the same extent as self-directed Google search. 
Taken together, our results suggest that people in the UK are increasingly turning to conversational AI for information about politics. These findings substantially extend prior work by demonstrating that conversational AI's effects on political knowledge generalise across multiple topics, political perspectives, and model families, suggesting that the shift toward AI-assisted political information-seeking may not lead to increased public belief in political misinformation.
\end{abstract}

\begin{significance}
As conversational AI systems rapidly replace traditional search engines,
concerns have emerged about their potential to spread political
misinformation and bias electoral outcomes. This study provides a broad epistemic health assessment in the context of conversational AI's actual use and impact on political knowledge. Using a representative UK survey during the 2024 election, we reveal widespread adoption: 13\% of eligible voters used
AI chatbots for election-related information-seeking. Through randomised
controlled trials, we demonstrate that conversational AI strengthens
belief in true information while reducing belief in misinformation as
effectively as traditional Google search, with effects generalising across
topics and models. 
While both AI and search guided
participants toward greater agreement with progressive-leaning factual
statements, this shift was comparable across the ideological spectrum. These findings suggest the transition to AI-assisted
political information-seeking may not exacerbate misinformation problems,
informing ongoing debates about AI governance and democratic integrity.
\end{significance}

\section*{Introduction}
Conversational AI systems (or chatbots) such as ChatGPT, Claude and Gemini are now used regularly by hundreds of millions of people across the world. Analysis of consumer usage reveals that among the most common use cases is the seeking of information for private and professional purposes, including both general knowledge and practical advice \citep{anthropic2024clio}. As public usage of chatbots continues to increase, a particular concern is that LLMs will produce unreliable or biased answers when queried about current affairs or political issues. This could be damaging to democracy, which relies on the electorate having access to reliable information, especially before elections \citep{summerfield2024democracy}. 

The concerns about AI's potential impact on political
knowledge are situated within a longstanding debate in political science
about whether factual information actually matters for democratic
decision-making. Some scholars have argued that voters can effectively
compensate for being relatively poorly informed by relying on heuristics, party affiliation, and opinion leaders \citep{lupia1994shortcuts, bartels1996uninformed}, suggesting that the quality of political information sources may be less consequential than commonly assumed. However, this optimistic view depends on the assumption that voters \textit{know} they are uninformed and respond accordingly. \citet{kuklinski2000misinformation} have shown that misinformed citizens, those who hold confident but incorrect
beliefs, represent a qualitatively different challenge: they make
systematically different policy judgements than both informed and
uninformed citizens, and their false confidence makes them more resistant to
correction of these beliefs. Indeed, attempts to correct political misinformation can fail or even backfire among ideologically committed individuals
\citep{nyhan2010corrections}, though more recent work suggests that such
backlash effects may be less prevalent than initially thought
\citep{guess2020backlash}. The broader information environment compounds these challenges: the rapid acceleration of content production and consumption has intensified competition for limited collective attention \citep{lorenzspreen2019accelerating}, while exposure to untrustworthy sources, though perhaps less prevalent than commonly feared \citep{guess2020exposure}, remains a persistent concern, particularly as social media has become a major outlet for political information, with false news stories reaching wide audiences across the ideological spectrum \citep{allcott2017social}. Moreover, exposure to opposing viewpoints does not straightforwardly reduce polarisation, but can rather reinforce pre-existing attitudes \citep{bail2018exposure}.
Against this backdrop, the emergence of conversational AI as a new source of political information raises the question of whether these systems will exacerbate or ameliorate existing challenges to an informed electorate.%}

Two main uncertainties make the extent of  potential risks unclear. First, it is unclear whether people trust AI enough to use it for political information-seeking in high-stakes political moments, such as in the lead-up to national elections. Among the general public, trust in AI is low: surveys consistently report that a majority of people do not trust AI to be used as tools to provide medical care, legal advice, or to assist human journalists \citep{newman2024digital, gillespie2023trust, mcclain2024chatgpt}. One recent survey showed that levels of trust in AI were comparable to those for politicians, advertising executives, and social media influencers who are themselves the least trusted of all professions \citep{laher2024trust}. Given this, it is an open question whether people will turn to conversational AI for political guidance during elections, even when such tools are available. 

Second, it is unclear whether conversational AI systems present functional substitutes for traditional search engines on common political issues. On the one hand, it is widely acknowledged that AI models have a problem with factuality (AI developers have devoted considerable energy to studying what they call ``hallucinations'' in AI models \citep{huang2024hallucination, ji2023hallucination, augenstein2024factuality, wang2023factuality}) and there has been concern that models may be politically biased, especially in a progressive or libertarian direction \citep{santurkar2023opinions, hartmann2023ideology, rottger2024compass}. On the other hand, significant progress has been made in tackling these risks: developers have implemented techniques such as retrieval-augmented generation (RAG) \citep{lewis2020retrieval}, knowledge-graphs \citep{agrawal-etal-2024-knowledge}, fine-tuning on factuality rankings \citep{tian2024finetuning}, and semantic entropy methods for detecting confabulations \citep{farquhar2024detecting}. As a result, it remains unclear if reliability issues would substantively impact users’ political knowledge in a real-world information-seeking setting. 

%\textcolor{cyan}{
A growing literature has begun to characterise how conversational AI differs from traditional internet search as an information source, with implications that cut both ways. On the demand side, evidence suggests that users already blend AI assistants into their information-seeking rather than substituting them fully, often acting on AI-provided information even when they doubt it and intending to verify it later \citep{wardle2025evolving}. On the supply side, experimental work shows that LLM-based search can speed up decisions at accuracy comparable to conventional search, but can also induce overreliance when the model is wrong, an effect attenuated when supporting evidence or model confidence is made visible \citep{spatharioti2025effects}. Relatedly, identical content is judged more credible when delivered in a conversational format than as static, search-style text, with users correspondingly less likely to catch inaccuracies \citep{anderl2024conversational}. Together, these findings motivate a direct test of whether conversational AI serves as a knowledge-equivalent substitute for internet search on common political issues, rather than merely a faster or more persuasive, but potentially less scrutinised interface.%}

A growing body of work has examined AI's potential to
influence political attitudes, demonstrating that AI-generated messages
can persuade humans on policy issues
\citep{bai2025llm, hackenburg2024evaluating} and that AI-generated
propaganda can be as persuasive as human-written content
\citep{goldstein2024how}. Recent work has further shown that the
persuasive power of conversational AI stems primarily from
post-training and prompting techniques rather than model scale or
personalisation, though with a concerning trade-off: optimising AI
systems for persuasion systematically reduces the factual accuracy of
their outputs \citep{hackenburg2025levers}. However, conversational
AI's persuasive capabilities may also be leveraged for positive ends:
\citet{costello2024durably} demonstrated that personalised dialogues
with an LLM durably reduced conspiracy beliefs, with effects
persisting for at least two months, suggesting that AI's capacity to
generate tailored, evidence-based arguments can help correct
misinformation rather than spread it. However, these studies examine scenarios in which AI-generated content is delivered to users, rather than the common scenario in
which users actively seek out political information from AI systems.%} 
Recent work  has begun to unravel the role of conversational AI in this latter use case \citep{taylor2024ai}, providing initial evidence that conversational AI can improve factual knowledge on specific political
topics. However, this preliminary work was limited to single-issue investigations favouring liberal-aligned factual information with a narrow scope of political issues, without built-in controls for general knowledge acquisition or examination of effects across the political spectrum.

In the present investigation, we first show in a representative survey %\textcolor{cyan}{
(Study 1)%} 
one week after the UK 2024 general election that as many as 13\% of eligible voters may have used conversational AI to find information relevant to their electoral choice. Second, in a randomised controlled trial (RCT, %\textcolor{cyan}{
Study 2%}
) we find that participants who used a chatbot (GPT-4, Claude, or Mistral) to research factual information related to issues of concern for UK voters increased their political knowledge to the same extent as participants who researched the same issues using Google search. In a follow-up RCT %\textcolor{cyan}{
Study 3)%}
, we find that even when LLMs were explicitly prompted to use sycophantic or persuasive techniques, participants' belief formation did not differ from those interacting with standard unprompted AI models. Taken together, these results suggest that although people in the UK are increasingly turning to AI for political information, this shift may not lead to increased public belief in political misinformation.

Our work makes three key contributions: (1) Replication: We replicate recent findings that conversational AI can improve political knowledge on specific issues \citep{taylor2024ai}; (2) Generalisation: We demonstrate that these effects generalise across multiple political topics, model families, prompting strategies, and remain consistent across politically balanced information representing both progressive and conservative viewpoints; (3) Broader epistemic health assessment: We provide the first assessment of AI's effects on broader epistemic health indicators (trust, private beliefs, extremity) and the first nationally representative data for the UK on real-world chatbot usage for political information during a national election.
Importantly, our work examines a specific but common
use case: structured, task-directed political information-seeking. Real-world interactions with conversational AI are more varied: people may use chatbots to seek validation for existing views, to
discuss politics in less structured ways, or may be redirected away from political engagement entirely. Our findings speak to the direct effects of using AI as a research tool for political information, but
should not be interpreted as capturing the full equilibrium effects of AI on political knowledge or engagement more broadly.%}

\begin{figure*}
    \centering
    \includegraphics[width=\linewidth]{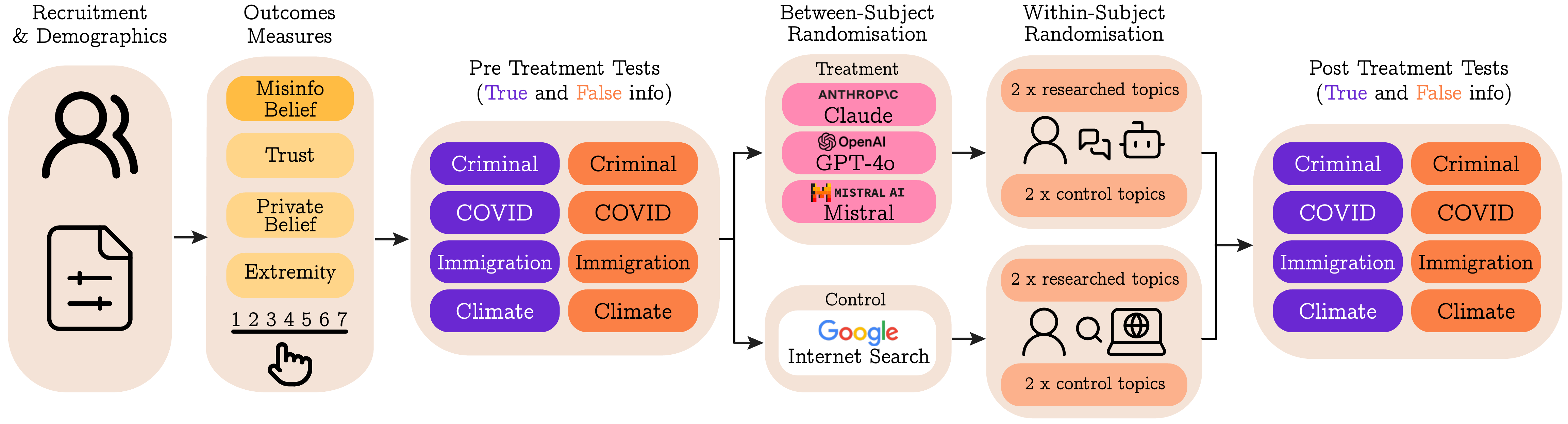}
    \caption{Experimental design for measuring the impact of conversational AI on political knowledge. Participants completed baseline assessments of misinformation belief (primary outcome) across four political topics (criminal justice, COVID-19, immigration, and climate change) using 7-point Likert scales. To assess generalization to broader measures of epistemic health, participants also completed assessments of trust levels, private political beliefs, and extremity indicators. Participants were randomised to using conversational AI chatbots (Claude, GPT-4, or Mistral) or Google search. During the research phase, participants investigated two randomly assigned topics (researched topics) while two others serve as within-subject controls (non-researched topics). Following the research phase, all measures were re-administered to assess pre-post changes.}
    \label{fig:explainer-fig}
\end{figure*}

\section*{Results}

\subsection*{Survey %\textcolor{cyan}{
(Study 1)}%} 
\begin{figure*}
    \centering
    \includegraphics[width=\linewidth]{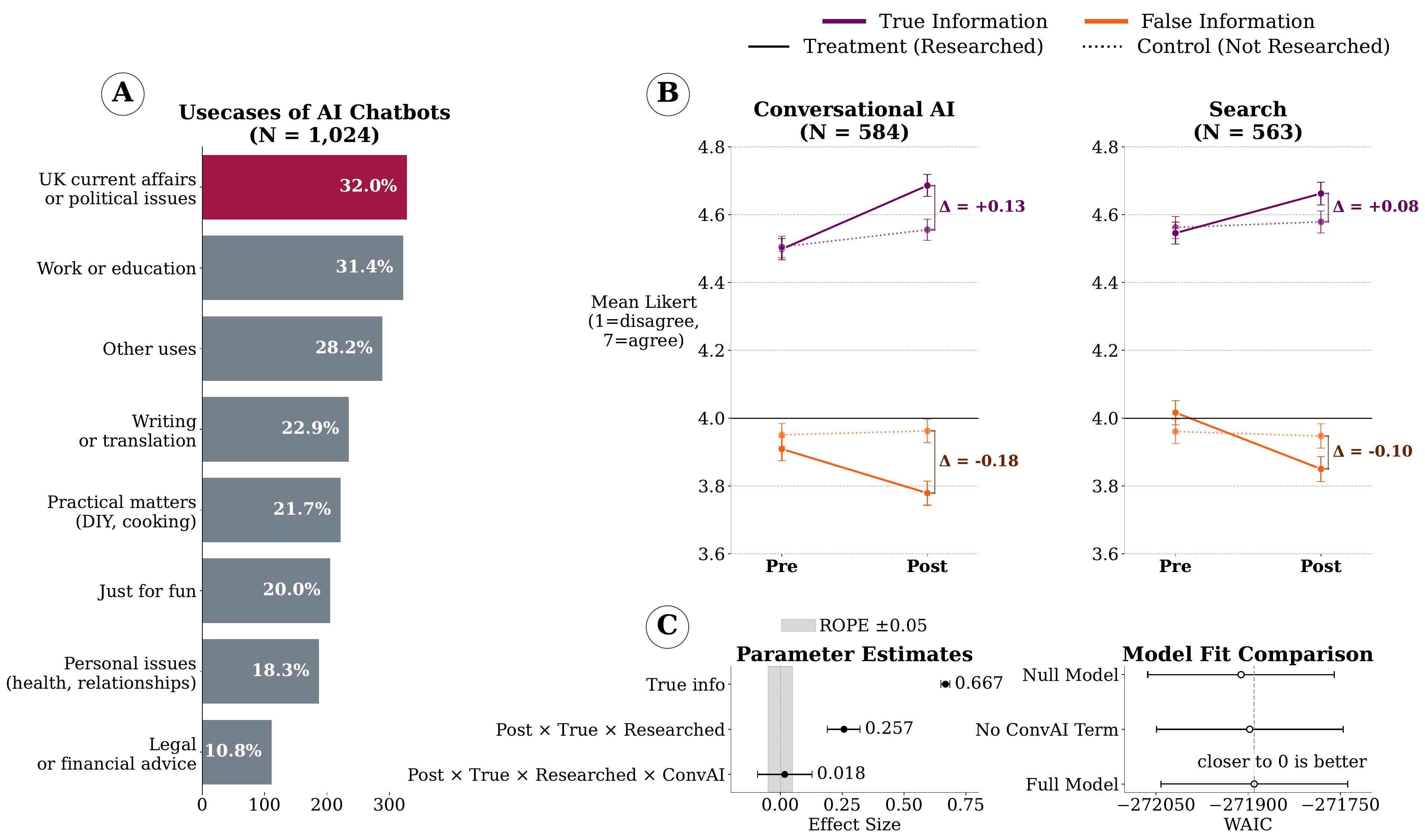}
    \caption{Conversational AI usage patterns and influence on belief in true versus false information. (A) Survey %\textcolor{cyan}{
    (Study 1)%} 
    results: Self-reported use cases for AI chatbots among UK users. (B) RCT %\textcolor{cyan}{
    (Study 2)%}
    results: Change in agreement with true (purple) vs. false information (orange) from pre to post researching. Left panel shows the conversational AI condition; right panel shows the Google search control condition. Solid lines indicate researched topics, dotted lines denote non-researched topics. Error bars represent 95\% Confidence Intervals. (C) RCT %\textcolor{cyan}{
    (Study 2)%} 
    results, left: Bayesian GLM parameter estimates, error bars denote Highest Posterior Density Interval (HPDI). Gray shaded area depicts an apriori defined region of practical equivalence (ROPE), where effect sizes are considered to be negligible/practically 0; Right: GLM comparison using Widely Applicable Information Criterion (WAIC), as a measure of out-of-sample predictive accuracy of the GLMs (closer to 0 is better). Full model = GLM1: GLM including parameters to quantify differences in change effects between conversational AI and Google search conditions, No ConvAI Term = GLM2: GLM not including parameters to quantify differences between different conversational AI models, Null model = GLM3: GLM not including parameters to quantify differences in change effects between conversational AI and Google search conditions or different conversational AI models}
    \label{fig:main-fig}
\end{figure*}

A representative sample of UK adults eligible
to vote ($N$ = 2,499) were surveyed in the four days immediately
following the UK general election that took place on July 4th 2024
(see Supplementary Information for details on survey demographics
and survey questions). Respondents primarily relied on traditional
media for political information over the preceding four weeks:
television remained the most common source (54\%), followed by
social media (36\%), internet websites (33\%), internet search
(29\%), radio (28\%) and newspapers (27\%). By comparison, 9\%
used AI chatbots as a source of political information. Among
chatbot users
($N$ = 1,024)%}
, our central finding
is that one-third (32\%) used chatbots in the lead up to the UK
2024 election to research information relating to current affairs
and political issues (Fig. \ref{fig:main-fig}A). This comprises
the most popular use case, on par with work or educational uses
(McNemar's $\chi^2(1)=0.1, p = 0.753$).

Participants who used chatbots for political
information%}
found them significantly more useful
than non-useful (89\% vs 11\%,
$N$ = 430; %} 
$\chi^2(1) = 256.3, p < .001$) and more accurate than inaccurate
(87\% vs 13\%, 
$N$ = 417;%} 
$\chi^2(1)=223.1, p < .001$). Most respondents viewed chatbots as
politically neutral rather than showing partisan bias (62\% vs
38\%, 
$N$ = 404;%} 
$\chi^2(1)=24.8, p < .001$).
Among those perceiving bias ($N$ = 152)%}
, there
was an equal split between right- and left-leaning ideologies
(58\% vs 42\%, $\chi^2(1)=3.8, p=0.052$). While respondents were
evenly divided on whether chatbots influenced their perspective
overall (47\% vs 53\%, 
$N$ = 422; %}
$\chi^2(1)=1.9, p = .173$),
among those reporting a directional influence ($N$ = 124)%}
, liberal influence
significantly exceeded conservative influence (63\% vs 37\%,
$\chi^2(1)=8.3, p = .004$). The majority of respondents felt no
influence on their voting intentions (60\% vs 40\%,
$N$ = 443; %} 
$\chi^2(1)=18.7, p < .001$), but
among those reporting an influence ($N$ = 176)%}
, most were encouraged rather than discouraged to vote (79\% vs
21\%, $\chi^2(1)=59.1, p < .001$).

\subsection*{Randomised Controlled Trials (RCT, Study 2 and 3) %\textcolor{cyan}{Study 2 and 3})
}
In Study 2 (RCT) %\textcolor{cyan}{Study 2 (RCT)%}
, we recruited an independent sample of UK residents (\textit{N} = 1,147 final sample) online via Prolific.com. Participants were then linked to a custom-built online experiment app and instructed to research true or false information related to issues of concern for UK voters (each participant was pseudo-randomly assigned to researching two out of four topics: climate change, immigration, criminal justice, COVID-19 policy; see Supplementary Information for details on sourcing and political balancing of the material). The two other topics served as within-subject controls (non-researched topics) to rule out non-specific temporal changes to beliefs.

Participants researched the two assigned topics consecutively, either using conversational AI (GPT-4o, Claude-3.5, or Mistral) or Google search (Fig. \ref{fig:explainer-fig}). Both setups were embedded within the online experiment app -- participants interacted with conversational AI through an embedded chat window, and Google search was available as an embedded browser window within the app.  The experiment app also recorded the time participants spent researching each topic.

We assessed beliefs in true and false information on a 7-point Likert scale before and after researching all topics. Since participants answered questions about all four topics both before and after the research phase, regardless of whether they researched them, we could compare belief change on topics participants actively researched against belief change on topics they were merely re-assessed on, isolating the effect of the research activity from generic effects of questionnaire repetition or participant engagement. Additionally, we measured other markers of epistemic health as secondary outcomes: trust, private political beliefs, and extremity change. After completing the study, participants were fully debriefed about the aims and hypotheses of the research.

To test our research questions, we fitted and compared three Bayesian Generalised Linear Models (GLM1-3, Eq. 1, see Methods for details). There was no model evidence of differences between conversational AI and Google search conditions. 

\begin{figure*}
    \centering
    \includegraphics[width=\linewidth]{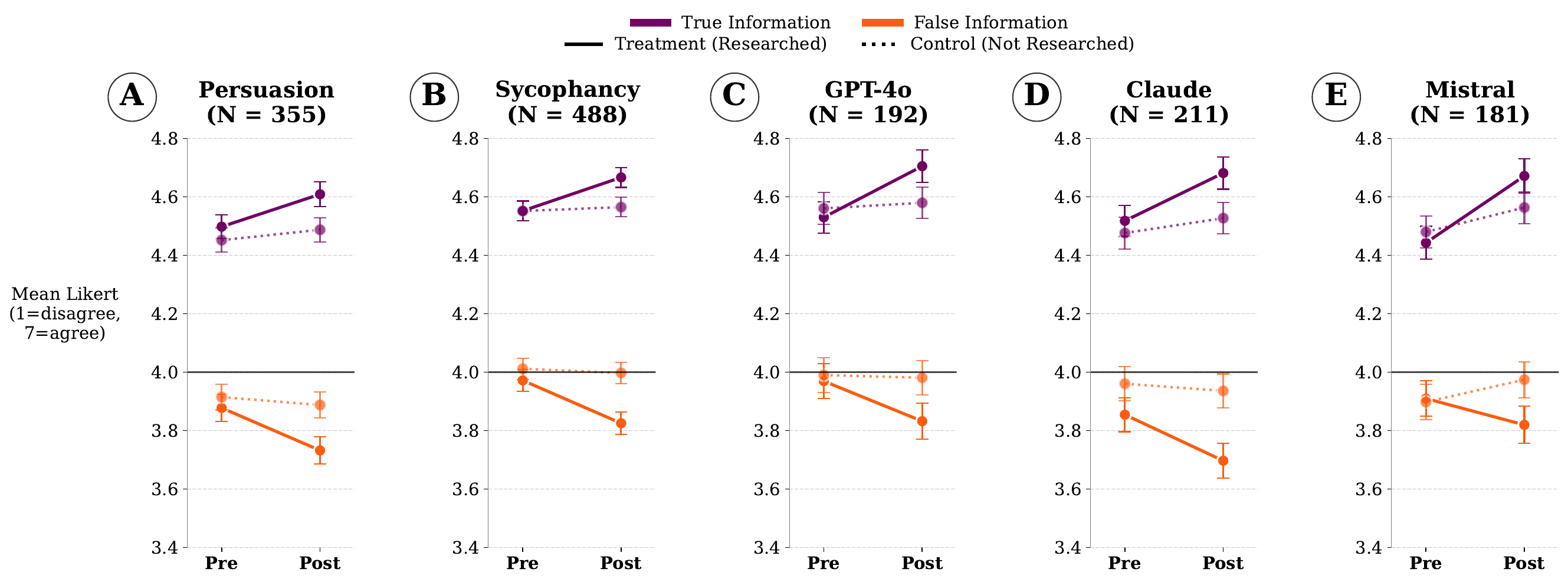}
    \caption{Belief in true and false information across prompting techniques %\textcolor{cyan}{
    (Study 3)%}
    and different conversational AI models %\textcolor{cyan}{
    (Study 2)%}
    . Change in agreement with true (purple) vs. false information (orange) from pre to post researching for: (A) GPT-4o prompted to be persuasive, (B) GPT-4o prompted to be sycophantic, (C) GPT-4o with standard prompting, (D) Claude, and (E) Mistral. Solid lines indicate researched topics, dotted lines denote non-researched topics. Error bars represent 95\% Confidence Intervals.}
    \label{fig:persuasion-sycophancy-models-plot}
\end{figure*}

Model fit metrics suggested no better fit of GLMs that included parameters for differences in change effects between conversational AI and Google search conditions (GLM1: Full model) in comparison to a GLM that did not include these terms (GLM3: Null model) (Fig. \ref{fig:main-fig}C, right panel). True information received on average approximately one Likert scale point higher agreement ratings than false information (Fig. \ref{fig:main-fig}B, purple vs orange lines), resulting in a non-zero difference parameter in GLM2 ($\beta_{\mathrm{TRUE}} = .67$, 95\%-Highest Posterior Density Interval (HPDI) [.65; .68] (Fig. \ref{fig:main-fig}C right panel). Researching (vs not researching) political issues increased belief in true information and decreased belief in misinformation across time points (Fig. \ref{fig:main-fig}B, solid vs dotted lines; ($\beta_{\mathrm{POST}\times\mathrm{TRUE}\times\mathrm{RESEARCHED}} = .26$, 95\%-HPDI [.19; .32], Fig. \ref{fig:main-fig}C). Importantly, we found that belief change was nearly identical for participants who researched using conversational AI or Google search (Fig. \ref{fig:main-fig}B, left vs right panel), reflecting in a close to zero parameter estimate ($\beta_{\mathrm{POST}\times\mathrm{TRUE}\times\mathrm{RESEARCHED}\times\mathrm{CONVAI}} = .02$, 95\%-HPDI [--.09; .12], Fig. \ref{fig:main-fig}C). This null difference between conditions was further qualified by the fact that the HPDI fully encloses a region of practical equivalence, (ROPE; gray shaded zone), an apriori specified interval of effect sizes that are negligible.
This pattern also held separately for each of the different model families tested (GPT, Claude, Mistral, \ref{fig:persuasion-sycophancy-models-plot}C-E).

\begin{figure*}
    \centering
\includegraphics[width=\linewidth]{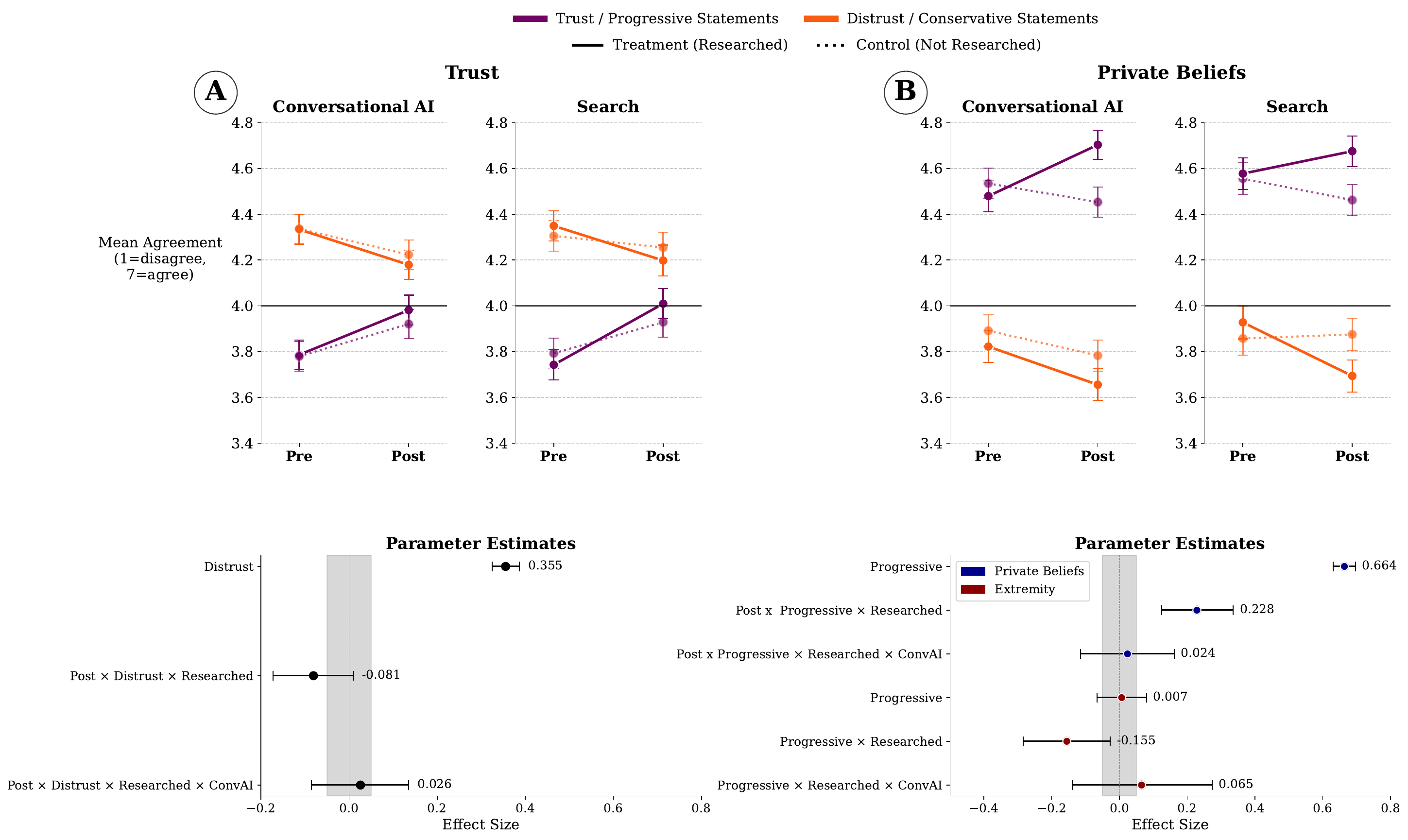}
    \caption{Agreement with trust and distrust statements and private beliefs. Top row: (A) Change in agreement with trust (purple) vs. distrust statements (orange) from pre to post researching, (B) change in private beliefs: agreement with progressive (purple) and conservative statements (orange) from pre to post researching. Error bars in top row represent 95\% Confidence Intervals. Bottom row:  Bayesian GLM parameter estimates for (A) agreement with trust/distrust statements and  (B) agreement with private belief statements (blue dots). Extremity was computed based on pre to post researching change in the sign of the difference to the center point of the Likert scale (4), indicating a flip on more agreement with progressive to more agreement with conservative beliefs (or vice versa) (red dots). Error bars denote Highest Posterior Density Interval (HPDI). Gray shaded area depicts an apriori defined region of practical equivalence (ROPE), where effect sizes are considered to be negligible/practically 0. Solid lines indicate researched topics, dotted lines denote non-researched topics.}
    \label{fig:trust-private-beliefs-plot}
\end{figure*}

We repeated the above analyses for trust, private political beliefs and extremity change, and found highly similar results as for beliefs in true and false information (Fig. \ref{fig:trust-private-beliefs-plot}). Trust was measured by asking participants to state their agreement with statements on trust or distrust in politicians, experts, media and technology (see Supplementary Information for details). 
We found that distrust statements on average produced higher agreement than trust statements ($\beta_{DISTRUST} = .35$, 95\%-HPDI [.32; .39]). There was no difference in change of trust ratings from before to after researching for researched vs not researched topics ($\beta_{POSTxDISTRUSTxRESEARCHED} = -.08$, 95\%-HPDI [--.17; .01]). Importantly, we found that the patterns of trust and trust change were highly similar for participants who researched using conversational AI or Google search ($\beta_{POSTxDISTRUSTxRESEARCHEDxCONVAI} = .026$, 95\%-HPDI [--.09; .14], Fig \ref{fig:trust-private-beliefs-plot}A). 
Refitting the best-fitting GLM excluding the items on trust in politicians (which had slightly different framing than other items) yielded qualitatively identical results. All reported non-zero effects remained non-zero and all null effects remained null. Specifically, the effect of distrust remained robust ($\beta_{DISTRUST} = 0.12$, 95\%-HPDI [0.08; 0.15]; 0\% ROPE overlap). There was still no difference in change of trust ratings from before to after researching for researched vs not researched topics ($\beta_{POSTxDISTRUSTxRESEARCHED} = -.07$, 95\%-HPDI [–0.15; 0.01]), and crucially, the patterns of trust and trust change remained highly similar for participants who researched using conversational AI or Google search ($\beta_{POSTxDISTRUSTxRESEARCHEDxCONVAI} = .02$, 95\%-HPDI [–0.07; 0.11]). This confirms that the inconsistent phrasing of this single item did not meaningfully affect our conclusions.%}

Private political beliefs were assessed by asking participants to state their agreement with progressive and conservative views on the 4 topics (see Supplementary Information for details). We observed that progressive statements on average produced higher agreement ratings than conservative statements  ($\beta_{PROGRESSIVE} = .66$, 95\%-HPDI [.63; .70]). Additionally, researching topics increased agreement with progressive views and decreased agreement with conservative views ($\beta_{POSTxPROGRESSIVExRESEARCHED} = .23$, 95\%-HPDI [.12; .34]). Importantly, we found that the effect of researching topics on private political beliefs was highly similar for participants who researched using conversational AI or Google search ($\beta_{POSTxPROGRESSIVExRESEARCHEDxCONVAI} = .02$, 95\%-HPDI [--.11; .15], Fig. \ref{fig:trust-private-beliefs-plot}B).

%{\color{cyan} 
To investigate whether the observed progressive shift is an artifact of our predominantly progressive-voting sample (progressive: 50.1\%; conservative: 23.7\%; no clear self-placement: 26.1\%), we added a binary ideological self-placement variable (progressive: Labour, Green, Liberal Democrats, SNP, Sinn Féin, Plaid Cymru; conservative: Conservative, Reform UK, Unionist parties; based on self-identification) and its interactions with all experimental predictors. Model comparison confirmed ideological self-placement as a strong predictor of beliefs overall ($\Delta$WAIC = 1,266), driven by ``partisan congruence'', i.e., participants agreed more with statements aligned with their political identity ($\beta_{PROGRESSIVExVOTE} = 2.00$, 95\%-HPDI [1.91; 2.08]). Critically, however, none of other higher-order interactions involving both pre-post treatment and ideological self-placement were meaningfully different from zero, including the key $\beta_{POSTxPROGRESSIVExRESEARCHEDxVOTE} = .04$ (95\%-HPDI [--.17; .26] and $\beta_{POSTxVOTE} = .003$, (95\%-HPDI [--.08; .08]), confirming that the progressive shift was comparable across the ideological spectrum. In other words, ideological self-placement (unsurprisingly) strongly predicts \textit{what} people believe on average, but not \textit{how they change} in response to researching information. Both conservative and progressive participants experienced a similar progressive shift, suggesting a general effect rather than a sample composition artifact. Additionally, we observed that exposure to information sources was linked to reduced partisan congruence ($\beta_{PROGRESSIVExRESEARCHEDxVOTE} = -.30$, 95\%-HPDI [--.44; --.16]). However, we interpret this finding with caution, as the observed effect is not a temporally specific change but rather represents an average effect over time.%}

Extremity change was defined as sign flips in private political beliefs from before to after researching/not researching an issue -- with reference to the center point of the Likert scale (4, values below this value being negative, and values above being positive). Within a Binomial GLM, there was no difference in extremity change for progressive or conservative statements ($\beta_{PROGRESSIVE} = .007$, 95\%-HPDI [--.07; .08]). Additionally, there was no reliable evidence that researching topics changed extremity of views differentially for progressive or conservative information ($\beta_{PROGRESSIVExRESEARCHED} = -.16$, 95\%-HPDI [--.28; --.03]; non-zero, but overlapping with ROPE). Again, we found that the effect of researching topics was nearly identical for participants who researched using conversational AI or Google search ($\beta_{PROGRESSIVExRESEARCHEDxCONVAI} = .07$, 95\%-HPDI [--.14; .27], Fig. \ref{fig:trust-private-beliefs-plot}B, bottom).

The above results were obtained using LLMs with standard prompts. However, models could be prompted to behave in ways that bias the user towards one view or another (persuasion), or to behave ``sycophantically'', refusing to contradict the user and reinforcing their existing beliefs. Next, thus, we asked whether AI systems prompted in this way might influence political informedness, belief, or trust to a greater extent than search engines. 

In %\textcolor{cyan}{
Study 3%}
, a structurally similar RCT to %\textcolor{cyan}{
Study 2%} 
(\textit{N} = 1,711 final sample), we investigated how a chatbot (GPT-4o) prompted to be sycophantic or persuasive affect beliefs in true and false information (and secondary outcomes) relative to an unprompted baseline GPT-4o (baseline/control condition). Chatbot system prompts were inserted programmatically via the app before the participant's first message, without the participant's knowledge (see Supplementary Information for exact prompts). The sycophantic system prompt instructed the LLM to support the users' pre-existing beliefs on the issue, irrespective of whether they agreed or disagreed with the issue (see Supplementary Information Box 1). Similarly, the persuasive system prompt instructed the LLM to support a randomly chosen view points (agree/disagree, which corresponded to the users' pre-existing beliefs in 50\% of the cases, see Supplementary Information Box 2).

The GLM comparison results and parameter estimates obtained were similar to the previous results with standard prompt settings (Fig. \ref{fig:persuasion-sycophancy-models-plot}A-B); no differences were found between prompted vs. unprompted LLMs $\beta_{\mathrm{POST}\times\mathrm{TRUE}\times\mathrm{RESEARCHED}\times\mathrm{PROMPT}} = .03$, 95\%-HPDI [--.06; .13] -- suggesting that interacting with LLMs prompted to be sycophantic or persuasive did not change participants views above and beyond the view changes achieved by baseline conversational AI models (and by extension, traditional Google search).

Participants' debrief ratings of model reliability
and agreement did not differ between prompted and unprompted conditions (sycophancy: all $t \leq 0.51$, $p \geq .609$; persuasion: all $t \leq 1.61$, $p \geq .108$), suggesting that
participants perceived the prompted and unprompted models similarly.
It is possible that our sycophancy and persuasion manipulations did
not alter model behaviour as strongly as intended: for sycophancy,
recent work suggests that question-based interactions substantially
attenuate sycophantic behaviour in LLMs even when system prompts
instruct otherwise \citep{dubois2026askdonttellreducing}, and participants in our study likely engaged with the chatbot primarily
by asking questions. For persuasion, the system prompt explicitly
constrained the model to use only factual information and logical
arguments on well-covered political topics, which may have limited
the model's ability to provide unreliable information.%}

Even though we found no differences of researching political issues using conversational AI or Google search on epistemic health, we additionally investigated potential time efficiency effects of the search methods. We found that the use of conversational AI reduced the information procurement time by 6-10\% in comparison to self-guided Google search (average time spent on both research tasks ($\pm$ standard deviation) in minutes: 17.94 ($\pm$8.64) for conversational AI vs 19.82 ($\pm$8.91) for Google search; $\beta_{\mathrm{CONVAI}} = -.11$, 95\%-HPDI [--.16; --.07], Gamma GLM, Eq. 3). 

Our results replicate and extend recent findings that conversational AI can improve factual political knowledge on specific issues \citep{taylor2024ai}. We substantially extend this work in several key ways. First, while prior work examined single issues in isolation, our within-subjects design demonstrates that these effects generalize across four politically salient topics simultaneously (climate change, immigration, criminal justice, COVID-19). Second, our inclusion of both progressive- and conservative-aligned true information (see Supplementary Information) demonstrates that AI's knowledge-enhancing effects are not limited to liberal-favoured facts. Third, our within-subject control topics (non-researched issues) allow us to rule out generic effects of questionnaire repetition or participant engagement, isolating AI-specific learning effects. Fourth, we extend beyond single knowledge questions to examine broader epistemic health indicators including trust, private political beliefs, and extremity -- none of which showed differential effects between AI and traditional search.

\section*{Discussion}
Replicating and extending recent work \citep{taylor2024ai}, we
demonstrate that conversational AI now sits alongside Google search as a commonly used source of information during high-stakes political moments like national elections.
Our experimental findings speak specifically to task-directed political
information-seeking: participants were instructed to research specific political topics using either conversational AI or Google search. This represents an important and increasingly common use case, but does not
capture the full range of ways in which people interact with AI in political contexts.%} 
Going beyond prior single-issue investigations, we show that across four salient issues, politically balanced true and false information, and three model families, researching with chatbots raised belief in true facts and lowered belief in false claims to the same extent as self-guided Google search, even when the chatbots were prompted to be persuasive or sycophantic. Our within-subject controls for non-researched topics and secondary outcomes (trust, private beliefs, extremity) further demonstrate that these effects represent genuine knowledge acquisition rather than generic engagement or politically-biased persuasion. These findings stand in contrast to the popular assumption that the use of chatbots for seeking information about news or current affairs may inherently erode political knowledge.

Our multi-topic examination has three main implications. First, we replicate prior work showing that AI can improve political knowledge, while demonstrating for the first time that this effect generalizes across multiple topics, political viewpoints, and is not limited to liberal-favored information. Second, our results suggest that the large body of prior work suggesting low levels of public trust in AI belies actual public usage for information-seeking: in fact, our data suggest that information-seeking is the most popular use case for chatbots among UK citizens (surpassing professional use, writing/translation, and practical advice), and that users found their chatbots to be useful, accurate, and unbiased. Third, our experiment suggests that contrary to widespread concern about AI reliability and hallucinations, models do not increase belief in false information when used to research current affairs and political issues across diverse topics and political perspectives.

One notable finding that warrants further investigation is the null effect of conversational AI on trust. Despite widespread concern that AI-generated content may erode public trust in institutions, experts, and media, we found no evidence that researching political information using chatbots affected participants' trust ratings, compared to using traditional Google search. The disconnect between low stated trust in AI and the absence of any measurable impact on broader institutional trust raises important questions for future research: Does familiarity with AI through direct use gradually shift trust perceptions? Might longer or repeated interactions with chatbots eventually influence trust in ways that a single research session cannot capture? And could the null effect reflect a ceiling or floor effect in trust attitudes that are deeply entrenched and resistant to short-term interventions? Longitudinal studies tracking trust dynamics over sustained AI use would be particularly valuable in addressing these questions.

Our findings also touch on longstanding debates about
the role of factual information in democratic decision-making. While
some scholars have argued that voters can compensate for relatively low levels of
information through heuristics and shortcuts \citep{lupia1994shortcuts}, our results are more consistent with
the view that the quality of information sources matters: both AI
and search engines measurably shifted participants' beliefs toward
factual accuracy on politically salient topics. This is particularly
relevant given evidence that misinformed citizens make systematically different policy judgements than merely uninformed
ones \citep{kuklinski2000misinformation}. However, whether these
informational gains translate into downstream changes in political
attitudes or behaviour remains an open question. Prior work suggests that the relationship between information exposure and political decision-making is complex: exposure to counter-attitudinal
information does not reliably change minds
\citep{bail2018exposure, nyhan2010corrections}, though backlash
effects may be rarer than commonly supposed \citep{guess2020backlash}. Future work should examine whether
AI-assisted improvements in factual political knowledge persist
over time and whether they influence subsequent political attitudes, voting behaviour, or broader civic engagement.%}

These conclusions are bounded by scope: we focused on one country, a single election cycle, four issues, and short interactive sessions with a small sample of models. This sample of models did not include models that have been shown to be highly controversial and biased towards specific ends of the political spectrum. In this sense, our results might underestimate the belief changing effects that more biased models may have. We tested the effects of using conversational AI for researching political information against a no-AI self-guided Google search condition. While this represents a valid control to rule out generic effects of self-selection and confirmation bias, most internet search engines as of today provide AI-enhanced features, like AI-generated summaries of the most relevant information across websites. Additionally, we did not study the newest generation of chatbots, which are typically internet search-enabled, providing users with most up-to-date information. These constraints might render our study more likely to reflect a comparison to traditional methods of information search. However, field data that link chatbot use to downstream attitudes and behaviour remain extremely sparse. We thus believe our results represent an important baseline to reference future developments in AI-guided search of political information and political belief change. 

Perhaps most importantly, our experimental design examines task-directed information-seeking, where participants were
instructed to research specific political topics. Real-world AI use is considerably more varied: people may use chatbots to seek validation for pre-existing views, to discuss politics in unstructured ways, or
conversely, AI may redirect users away from political engagement entirely by making non-political uses more compelling. Our findings therefore speak to the direct epistemic effects of using AI as a political research tool, but should not be interpreted as evidence about the broader equilibrium effects of AI availability on political knowledge or engagement. Future work should examine more naturalistic, self-directed interactions with AI in political contexts, including longitudinal designs that capture how AI use patterns evolve over time and how they interact with broader media consumption habits.

Our findings should also be interpreted in light of
prior work demonstrating that search engines themselves are not neutral
information intermediaries. \citet{epstein2015search} showed %\textcolor{cyan}{
in an experimental setup%} 
that biased search engine rankings can shift voting preferences of undecided voters by 20\% or more, largely without their awareness, %\textcolor{cyan}{
an effect they termed%} 
the Search Engine Manipulation Effect (SEME). More recently, \citet{aslett2024online} demonstrated that searching online
to evaluate specific misinformation claims can paradoxically increase belief in them, particularly when search engines return low-quality results, a mechanism they attribute to ``data voids'', or informational spaces dominated by unreliable sources. These findings stand in apparent tension with our results, which show that both conversational AI and Google search reduced belief in misinformation across topics. We believe these findings are complementary rather than contradictory, and the divergence is informative about the boundary conditions of search-based information-seeking. Several key design differences may account for the different outcomes. First, our participants engaged in broad, self directed research on political topics rather than evaluating specific pre-selected misinformation claims. This general information-seeking task is more likely to surface high-quality, mainstream sources, particularly for well-covered topics like climate change, immigration, criminal justice, and COVID-19 policy, reducing the likelihood of encountering
the data voids identified by \citet{aslett2024online}. 
%\textcolor{cyan}{
This interpretation aligns with trace-data studies of real-world search behaviour: exposure to partisan or unreliable content on Google is driven more by users' own selections than by algorithmic curation \citep{robertson2023users}, and engagement with unreliable sites occurs predominantly when users deliberately navigate to them rather than encountering them through general topical queries \citep{greene2024current}. Broad, topic-level information-seeking of the kind our participants performed is therefore comparatively unlikely to steer users toward low-quality sources.%}
Similarly,
\citet{aslett2024online} found that their search effect was concentrated among individuals for whom search engines returned lower-quality information, and was absent when search results contained only high-quality sources. Second, the topics in our study are among the most extensively covered political issues in the UK, with substantial high-quality information available from mainstream outlets, government sources, and established think tanks. This likely
provided a protective buffer against data voids. Third, and importantly, participants in our study conducted their Google searches through an embedded browser window within our experiment app, which used a clean session without access to participants' personal browsing
history, cookies, or prior search data. This means that the search results participants encountered were not personalised based on their individual browsing behaviour or ideological profile, effectively controlling for the kind of algorithmic personalisation that can exacerbate filter bubbles and data voids
\citep{epstein2015search, aslett2024online}. In real-world search,
where results are tailored to users' prior behaviour, the effects on political knowledge could differ from what we observe here. 
%\textcolor{cyan}{
Consistent with the view that search outputs are themselves shaped by context, the framing of results can vary systematically: \citet{berkebileweinberg2025internet} find that image-search depictions of climate change, one of the four topics in our study, differ across countries in ways that shape perception.%}
However, our findings cannot be assumed to generalise to more niche or
emerging political topics where reliable information may be scarce and low-quality sources may dominate search results, nor to
personalised search environments where algorithmic curation may direct users toward lower-quality information. The interaction between topic salience, information availability, search personalisation, and the effects of AI-assisted information-seeking represents an
important avenue for future research. Additionally, \citet{epstein2015search}'s finding that biased search rankings can
powerfully influence voter preferences without awareness raises a broader concern about algorithmic curation of political information, whether through search engines or conversational AI. While our study found no differential effects between AI and search on political knowledge, this does not rule out the possibility that either tool could be deliberately manipulated to bias users, nor that algorithmic dynamics might organically favour certain viewpoints. Our finding that both AI and search guided participants toward greater agreement with
progressive-leaning factual statements may partly reflect such dynamics in the underlying information landscape, rather than tool-specific bias.%}

Additionally, while our sycophancy and persuasion
RCTs found no differential effects of prompting on belief change,
participant debrief ratings suggest that the manipulations may not
have altered model behaviour as strongly as intended. For
sycophancy, this is consistent with recent evidence that
question-based interactions attenuate sycophantic behaviour in
LLMs \citep{dubois2026askdonttellreducing}. For persuasion, the
constraint to use only factual information on well-covered topics
may have limited the model's ability to deviate from its baseline
behaviour. These results should therefore be interpreted with
caution, and future work should examine whether stronger
manipulations, for example, using models fine-tuned for
sycophantic or persuasive behaviour, or testing on more niche
topics where the factual landscape is less well-established,
might yield different outcomes. Importantly, however, the paper's
central claim that conversational AI increases political knowledge
to the same extent as Google search rests on the comparison between
standard unprompted AI and search, and is not contingent on the
prompting results.

While our study replicates the core finding of \citet{taylor2024ai} regarding AI's capacity to improve factual knowledge, our multi-topic, politically-balanced design with built-in controls provides substantially stronger evidence for the generalisability and political neutrality of these effects. However, like prior work, we remain limited to examining factual questions with objectively verifiable answers, and future research should examine more subjective or contested political claims where the ``correct'' answer is less clear-cut.

Within the above limits, our results suggest that for everyday task-directed information-seeking, today’s chatbots may perform on par with self-directed Google search -- potentially with no cost to political knowledge.

\newpage
\section*{Methods}
\subsection*{Survey}

For the survey %\textcolor{cyan}{
(Study 1)%}
, we recruited UK residents (\textit{N} = 2,499) online. For data plotting and statistical analyses, we reweighted respondents based on official census stats concerning age, gender, ethnicity, region, and socio-economic grade in the UK to correct any imbalances between the survey sample and the population to ensure it is nationally representative. For statistical analysis, we employed $\chi^2$ tests for independence when comparing proportions between different response categories within single-choice questions (e.g., useful vs non-useful responses), while McNemar's test was used for multiple-selection questions where respondents could select more than one option, as this test accounts for the dependency between paired responses from the same individuals (e.g., comparing selection rates between use cases of LLMs in the last 4 weeks where respondents could choose multiple use cases).

\subsection*{RCT}
For the RCTs %\textcolor{cyan}{
(Study 2 and 3)%}
, we recruited a separate, independent sample of UK residents
(\textit{N} = 2,858 final sample) online via Prolific. 

We did not conduct a formal power analysis, but oriented on other studies on similar research
questions that used similar sample sizes.
The study was approved by an internal committee within the UK Department of Science, Innovation and Technology (DSIT) that was set up specifically to review human-participants research, employing a framework called Responsible Research Framework. Our study was assigned reference number: 00001. The board reviewed both research ethics and data protection issues, including whether a data protection impact assessment (DPIA) was required and approved the study. All research was conducted in accordance with the Declaration of Helsinki.%} 
After completing the study, participants were fully debriefed about the aims and hypotheses of the research.

After recruitment, participants provided informed consent before being
linked to a custom-built online experiment app. Each participant was
pseudo-randomly assigned to researching two out of four topics 
(climate change, immigration, criminal justice, COVID-19 policy) -- these served as researched topics. The two other topics served as within-subject
controls (non-researched topics) to rule out non-specific temporal changes to beliefs. Participants researched the two assigned topics consecutively. 
In %\textcolor{cyan}{
Study 2%}
, participants were randomly assigned to conduct research either interfacing with a conversational AI model (GPT-4o, Claude-3.5, or Mistral) or
Google search (control condition, Fig. \ref{fig:explainer-fig}). Both setups were
embedded within the online experiment app -- participants interacted with
conversational AI through an embedded chat window, and Google search was
available as an embedded browser window within the app. The experiment app
also recorded the time participants spent researching each topic.

In a separate but structurally similar RCT (%\textcolor{cyan}{
Study 3%}
), we randomly assigned
participants to conduct research using only one of the conversational AI models (GPT-4o)
that was either instructed with a default system prompt (baseline/control
condition) or with a system prompt instructing it to be persuasive or
sycophantic (treatment conditions). System prompts were inserted
programmatically via the experiment app before the participant's first message,
without the participant's knowledge (see Supplementary Information for
exact prompts). The sycophantic system prompt instructed the LLM to
support the users' pre-existing beliefs on the issue, irrespective of
whether they agreed or disagreed with the issue. Similarly, the persuasive
system prompt instructed the LLM to support a randomly chosen viewpoint
(agree/disagree, which corresponded to the users' pre-existing beliefs in
50\% of the cases).

In both %\textcolor{cyan}{
Study 2 and 3%}
, we measured the effect of using an AI model or Google search in
researching political issues across four different outcomes on a 7-point
Likert scale, ranging from disagree to agree: belief in true and false
information, trust, private political beliefs, and extremity change. For
beliefs in true and false information, participants stated their level of
agreement or disagreement with 16 statements. Of these statements, 8 were
true and 8 were false; true statements were drawn from policy reports
published by reputable UK think tanks with variable political orientations
(see Supplementary Information for detailed statements). Participants
stated their agreement with statements on trust or distrust in
institutions, experts, media, and technology (see Supplementary
Information for detailed statements). Participants also indicated their
private political beliefs, operationalised by agreement with progressive or conservative leaning statements for the 4 topics (see Supplementary Information
for detailed statements).

There was no indication of systematic differences in dropout rates (after
starting the study and providing informed consent) between the
Conversational AI group (5.86\%) and Search group (7.34\%, $Z = -1.65$,
$p = .099$, two-proportion z-test), nor was there evidence for attrition
rate differences between different Conversational AI models used for
research or different prompting techniques (sycophancy or persuasion) (all
$p \geq .152$, two-proportion z-tests).

To test for selective attrition, we additionally regressed attrition status on treatment assignment, pre-treatment covariates (age, gender, income, religion, education, political leaning,
disability status, mental health status, and chatbot use frequency), and all treatment $\times$ covariate interactions, pooling across all RCT samples. A joint F-test of the interaction terms was non-significant ($F(23, 2170) = 0.89$, $p = .613$), suggesting that attrition was not systematically related to the combination of treatment assignment and participant characteristics.%}

Additionally, we assessed participant engagement using
topic-specific compliance questions embedded throughout the study. For
each of the four topics, we designed 5 factual questions testing
whether participants had engaged with the research material (e.g.,
``What percentage of the UK's electricity was generated from renewable
sources in 2023?'' for climate change; see Supplementary Information
for the full list). Each participant answered 10 compliance questions
in total (5 per each of their two researched topics). There were no
significant differences in compliance rates between the conversational
AI and Google search conditions, either within individual models (all
$t \leq 1.56$, $p \geq .120$, two-sample t-tests) or aggregated
across all models ($M_{CONVAI/PROMPTED} = .540$, $M_{SEARCH/UNPROMPTED} = .543$, $t =
0.42$, $p = .678$), suggesting that participant engagement was
comparable across experimental conditions. 
%\textcolor{cyan}{
As a sensitivity check, we additionally refit all Bayesian GLMs including individual subject-level compliance scores (fraction of questions answered correctly) as a covariate, and compared them to the models reported above. WAIC-based model comparison was identical in every dataset and condition. Across the best-fitting models per outcome, the maximum change in any posterior mean was 0.077; no effect changed in interpretation. The compliance coefficient itself was consistently near zero ($\beta_{COMPLIANCE} < 0.02$, 100\% ROPE overlap), indicating that compliance was not meaningfully related to any outcome variable. Together, these analyses suggest that non-compliance neither differed systematically across conditions nor meaningfully influenced our results.%}

In the sycophancy and persuasion RCT (%\textcolor{cyan}{
Study 3%}
), we additionally assessed participants' perceptions of model behaviour at debrief by asking them to rate the model's reliability and the extent to which they agreed with the model's replies, both on a 7-point Likert scale. Participants' ratings of model reliability and agreement did not differ between the prompted and unprompted (control) conditions in either the sycophancy study (all $t \leq 0.51$, $p \geq .609$) or the persuasion
study (all $t \leq 1.61$, $p \geq .108$). All mean ratings were close to the midpoint of the scale (range: 3.72--4.02), suggesting that participants perceived the prompted and unprompted models similarly.
For the sycophancy condition, one plausible explanation is that
participants primarily engaged with the chatbot by asking questions, a mode of interaction that has been shown to substantially attenuate
sycophantic behaviour in LLMs even when system prompts instruct
otherwise \citep{dubois2026askdonttellreducing}. For the persuasion
condition, the system prompt explicitly constrained the model to use only factual information and logical arguments, and participants were
researching well-covered political topics where the factual landscape
is well-established, which may have limited the model's ability to provide unreliable responses. These results are
consistent with our main finding of no differential effects of
sycophantic or persuasive prompting on belief change.%}

\subsection*{Statistical Modeling}

Belief in true and false information (agreement/disagreement with a presented issue statement across researched and non-researched topics) served as our primary outcome of interest, the other variables were analyzed as secondary outcomes. Extremity change was defined as before to after researching sign flips in the difference between private political beliefs -- center point of the Likert scale (4).  

Issue agreement and extremity data before and after search/AI conversation were analyzed using three Bayesian multilevel GLMs \citep{luettgau2025hibayeshierarchicalbayesianmodeling, dubois2025skewedscorestatisticalframework}. We specified GLMs with ordered-logistic likelihood functions to model the ordinal categories of issue agreement responses. For extremity data, we defined GLMs with Binomial likelihood function. GLMs were fitted using sampling-based Bayesian inference using Numpyro \citep{Phan2019} for Markov Chain Monte Carlo sampling (using No-U-Turn-Sampler – NUTS – a variant of Hamiltonian Monte Carlo \citep{Hoffman2011}) to estimate the posterior distribution of linear model parameters. These parameters include different intercept and slope parameters that combine linearly to influence the likelihood of ordinal or Binomial responses.

Weakly informative prior probability distributions were specified for each model, as indicated in the model specifications. We drew 4 x 2000 samples from the posterior probability distributions (4 x 2000 warmup samples) across four independent Markov chains. The quality and reliability of the sampling process were evaluated using the Gelman-Rubin convergence diagnostic measure ($\hat{R}$)  and by visually inspecting the trace- and rank-plots of the Markov chains.

For all models fitted, for all sampled parameters there were no divergent transitions between Markov chains for any reported models.

For GLM comparisons and to identify the best-fitting GLM for the observed data, we used the Widely Applicable Information Criterion (WAIC \citep{WatanabeSWATANAB2010}). Parameter estimates were considered non-zero if the Highest Posterior Density Interval (HPDI) around the parameter did not contain zero. The HPDI was compared to a region of practical equivalence (ROPE), i.e., an interval of parameter values $[-0.05; 0.05]$ representing the null hypothesis of the parameter being equivalent to 0.

Specifically, we defined an ordered-logistic multilevel GLM for agreement rating data, Eq. \ref{eq:ordlog_model}

\begin{equation}
\begin{aligned}
y_i &\sim \text{OrderedLogistic}(\eta_i, \kappa) \\
\eta_i &= \text{intercept}_{\text{overall}} + \text{intercept}_{\text{subject}} + \boldsymbol{X}_i \cdot \boldsymbol{\beta_i} \\
\kappa &= \text{cutpoints} \\
\text{intercept}_{\text{overall}} &\sim \text{Normal}(0, 0.01) & \\
\text{intercept}_{\text{subject}} &= \text{intercept}_{\text{subject}_{\text{raw}}} \cdot \sigma_{\text{subject}} \\
\text{intercept}_{\text{subject}_{\text{raw}}} &\sim \text{Normal}(0, 0.01) & \\
\sigma_{\text{subject}} &\sim \text{HalfNormal}(0.01) \\
\beta_{{\text{i}}_{\text{raw}}} &\sim \text{Normal}(0, 0.1) \\
\beta_{{\text{i}}_{\text{scale}}} &\sim \text{HalfNormal}(0.1) & \\
\beta_i &= \beta_{{\text{i}}_{\text{raw}}} \cdot \beta_{{\text{i}}_{\text{scale}}} \\
\text{cutpoints} &\sim \text{TransformedDistribution} \left( \text{Dirichlet}(\boldsymbol{\alpha}), \text{SimplexToOrderedTransform}(\text{0}) \right)  \\
\alpha &= 1
\label{eq:ordlog_model}
\end{aligned}
\end{equation}

For extremity data, we defined a Binomial multilevel GLM, Eq. \ref{eq:binomial_glm}
\begin{equation}
\begin{aligned}
y_i &\sim \text{Binomial}(n_{\text{i}}, p_i) \\
\eta_i &= \text{intercept}_{\text{overall}} + \text{intercept}_{\text{subject}} + \boldsymbol{X}_i \cdot \boldsymbol{\beta_i} \\
p_i &= \text{sigmoid}(\eta_i) \\
\text{intercept}_{\text{overall}} &\sim \text{Normal}(0, 0.1) & \\
\text{intercept}_{\text{subject}} &= \text{intercept}_{\text{subject}_{\text{raw}}} \cdot \sigma_{\text{subject}} \\
\text{intercept}_{\text{subject}_{\text{raw}}} &\sim \text{Normal}(0, 0.1) & \\
\sigma_{\text{subject}} &\sim \text{HalfNormal}(0.1) \\
\beta_{{\text{i}}_{\text{raw}}} &\sim \text{Normal}(0, 0.1) \\
\beta_{{\text{i}}_{\text{scale}}} &\sim \text{HalfNormal}(0.1) & \\
\beta_i &= \beta_{{\text{i}}_{\text{raw}}} \cdot \beta_{{\text{i}}_{\text{scale}}} \\
\label{eq:binomial_glm}
\end{aligned}
\end{equation}

For both GLMs, different numbers and combinations of predictors were contained in the design matrix $\boldsymbol{X}$. GLM1 (full model) contained all experimental factors [Post (pre or post timepoint), True (information true or false), Researched (topic researched or not), convAI (LLM or Google search), LLM-type (GPT-4o, Claude or Mistral), their two-, three-, four- and five-way interaction effects].

GLM2 (no convAI terms model) contained all of the above predictors, except for LLM-type and the associated interaction effects.
GLM3 (null model) contained all of the above predictors, except for LLM-type and convAI, and the associated interaction effects.
These three GLMs represent different hypotheses about the data that allowed us to make inferences about the underlying data generating process.
In case of testing sycophancy and persuasion prompts, LLM-type represented the prompting strategy vs unprompted LLM.
In the GLMs above, we use effect coding (-0.5; 0.5) for binary experimental factors and contrast coding [0.5; -0.25; -0.25] for representing differences between different LLMs.

To model information procurement time, we specified a hierarchical/multilevel GLM with Gamma likelihood function (Eq. \ref{eq:gamma_glm})

\begin{equation}
\begin{aligned}
y_i &\sim \text{Gamma}(\alpha, \beta_i) \\
\beta_i &= \frac{\alpha}{\mu_i} \\
\mu_i &= \exp(\eta_i) \\
\eta_i &= \text{intercept}_{\text{overall}} + \beta_{\text{convAI}} \cdot \text{convAI} + \text{intercept}_{\text{subject}} \\
\text{intercept}_{\text{overall}} &\sim \text{Normal}(0, 1) \\
\text{intercept}_{\text{subject}} &= \text{intercept}_{\text{subject}_{\text{raw}}} \cdot \sigma_{\text{subject}} \\
\text{intercept}_{\text{subject}_{\text{raw}}} &\sim \text{Normal}(0, 1) \\
\beta_{\text{convAI}} &\sim \text{Normal}(0, 1) \\
\sigma_{\text{subject}} &\sim \text{Exponential}(1) \\
\alpha &\sim \text{Exponential}(1)
\label{eq:gamma_glm}
\end{aligned}
\end{equation}

%\textcolor{cyan}{
For each of the four topics, we designed 5 factual multiple-choice
questions testing whether participants had engaged with the research
material. Each participant answered 10 compliance questions in total
(5 per each of their two researched topics, see SI for details). Because subject-level compliance was measured post-treatment, it was not included as a covariate in any model. We conducted a sensitivity check to assess differential effects of subject compliance using GLMs including this covariate (a numerical score on compliance sanity check data).%}

\section*{Data Availability}
The survey data, experimental data, and all analysis code necessary to reproduce the findings of this study will be made publicly available upon publication on GitHub at \url{https://github.com/lenluettgau/dhi1-analysis} (\cite{luettgau2025dhi1}). 

\section*{Conflict of Interest Statement}
Saffron Huang is employed at Anthropic PBC, San Francisco, USA. All her work on this paper was conducted whilst employed at UK AISI. She did not contribute any work after her move to Anthropic.
All other authors declare no conflict of interests.

\newpage
\bibliographystyle{plainnat}
\bibliography{dhi_1}

\newpage
\newcommand{\customsection}[1]{%
    \newpage
    \vspace*{2cm}
    {\centering\Large\bfseries #1\par}
    \vspace{1cm}
    \vspace{1cm}
}

\customsection{Supplementary Information}

\section*{Survey %\textcolor{cyan}{
(Study 1)%}  
Demographics}
For each variable we show the percentages in the weighted sample with the raw percentages in parentheses. 
\begin{itemize}
    \item Gender:
    \begin{itemize}
        \item Male: 48.34\% (51.62\%)
        \item Female: 51.46\% (48.18\%)
    \end{itemize}
    \item Age group:
    \begin{itemize}
        \item 18 to 24: 11.60\% (12.61\%)
        \item 25 to 34: 14.01\% (19.37\%)
        \item 35 to 54: 37.41\% (34.73\%)
        \item 55 to 64: 14.33\% (14.17\%)
        \item 65+: 22.65\% (19.13\%)
    \end{itemize}
    \item Generation:
    \begin{itemize}
        \item Generation Z: 15.13\% (17.69\%)
        \item Millennials: 25.29\% (29.33\%)
        \item Generation X: 29.93\% (26.69\%)
        \item Baby Boomers: 27.81\% (24.97\%)
    \end{itemize}
    \item Region:
    \begin{itemize}
        \item London: 13.05\% (15.73\%)
        \item Rest of South: 31.57\% (28.13\%)
        \item Midlands: 16.05\% (16.81\%)
        \item North: 23.37\% (23.69\%)
        \item Wales: 4.72\% (5.20\%)
        \item Scotland: 8.36\% (7.96\%)
        \item Northern Ireland: 2.92\% (2.48\%)
    \end{itemize}
    \item Social Grade:
    \begin{itemize}
        \item ABC1: 56.22\% (60.10\%)
        \item C2DE: 42.74\% (38.86\%)
    \end{itemize}
    \item Work Status:
    \begin{itemize}
        \item Full time: 43.46\% (50.78\%)
        \item Part time: 16.89\% (15.57\%)
        \item Unemployed: 8.40\% (7.68\%)
        \item Retired: 21.33\% (17.65\%)
        \item Other: 9.40\% (7.96\%)
    \end{itemize}
    \item Working Status (Aggregated):
    \begin{itemize}
        \item Working (All): 60.34\% (66.35\%)
        \item Not Working (All): 39.14\% (33.29\%)
    \end{itemize}
    \item Annual Household Income:
    \begin{itemize}
        \item Under £14k: 13.77\% (12.48\%)
        \item £14k to £21k: 10.60\% (9.60\%)
        \item £21k to £34k: 27.05\% (24.73\%)
        \item £34k to £48k: 16.93\% (17.33\%)
        \item More than £48k: 25.89\% (31.41\%)
    \end{itemize}
\end{itemize}

\section*{Survey %\textcolor{cyan}{
(Study 1)%} 
Questions and Response Options}

\subsection*{Q1a}
Over the past four weeks, which of the following have you used to find out about UK political issues or current affairs?

\begin{itemize}[noitemsep]
\item Newspapers (print or website / app, for example Daily Mail or Guardian Online)
\item Television (including live streaming, on demand and broadcast)
\item Radio
\item AI chatbots (for example ChatGPT)
\item Social media sites (for example Facebook, Instagram, TikTok, Twitter/X, YouTube)
\item Other internet sites, (for example BBC News)
\item Podcasts
\item Internet Search (for example Google)
\item Other, please specify
\item None of the above
\end{itemize}

\subsection*{Q1b}
In general, how much do you trust the following sources to provide accurate information about UK political issues or current affairs?

\textbf{Sources evaluated:}
\begin{itemize}[noitemsep]
\item Newspapers (print or website / app, for example Daily Mail or Guardian Online)
\item Television (including live streaming, on demand and broadcast)
\item Radio
\item AI chatbots (for example ChatGPT)
\item Social media sites (for example Facebook, Instagram, TikTok, Twitter/X, YouTube)
\item Other internet sites, (for example BBC News)
\item Podcasts
\item Internet Search (for example Google)
\end{itemize}

\textbf{Response options for each source:}
\begin{itemize}[noitemsep]
\item Trust a great deal
\item Trust to some extent
\item Do not trust very much
\item Do not trust at all
\item Don't know
\end{itemize}

\subsection*{Q1c}
How often, if at all, do you use AI chatbots in your professional life or leisure time?

\begin{itemize}[noitemsep]
\item I have never used an AI chatbot
\item I use a chatbot from time to time
\item I use a chatbot around once a week
\item I use a chatbot almost every day
\item I use a chatbot at least once a day
\item I don't know
\end{itemize}

\subsection*{Q1d}
Which, if any, of the following AI chatbots have you used in the last four weeks?

\begin{itemize}[noitemsep]
\item ChatGPT
\item Gemini/Bard
\item Claude
\item Bing AI
\item Pi
\item Perplexity AI
\item Open source chatbots, for example Llama or Mistral
\item Other
\item None of these/ I haven't used AI chatbots in the last four weeks
\end{itemize}

\subsection*{Q1e}
In the last four weeks, have you asked an AI chatbot for information or advice on any of the following topics?

\begin{itemize}[noitemsep]
\item Personal issues, such as health or relationships
\item Practical matters, such as DIY or cooking
\item Legal or financial advice
\item Information to help me with my work or education
\item UK current affairs or political issues (including information about the UK general election)
\item Help with translation, or help composing a piece of writing
\item I tried to engage the chatbot in a conversation just for fun
\item I used a chatbot for something else
\end{itemize}

\subsection*{Q2a}
In the last four weeks, did you see social media content focussed on UK political issues that you suspected to be AI-generated?

\begin{itemize}[noitemsep]
\item Yes, but the content did not appear to be misleading
\item Yes, and the content appeared to be misleading
\item No, I have not seen social media posts that I suspected to be generated by AI
\item No, I am not on social media
\item I don't know
\end{itemize}

\subsection*{Q2b}
In the last four weeks, did you see news articles focussed on UK political issues that you suspected to be AI-generated?

\begin{itemize}[noitemsep]
\item Yes, but the content did not appear to be misleading
\item Yes, and they appeared to contain misleading content
\item No, I have not come across online news articles that I suspected to be generated by AI
\item No, I do not read news online
\item I don't know
\end{itemize}

\subsection*{Q2c}
In the last four weeks did you see online images or videos on any subject that you suspected to be AI-generated fakes (often called deepfakes)?

\begin{itemize}[noitemsep]
\item Yes, I have come across images or videos that I suspected to be deepfakes (other than those that were labelled as such for reporting purposes, for example fact-checking)
\item No, I have not seen images or videos that I suspected to be deepfakes
\item I don't know
\end{itemize}

\subsection*{Q3a}
In the last four weeks, have you seen information about UK current affairs or political issues from any of the following sources?

\textbf{Sources evaluated:}
\begin{itemize}[noitemsep]
\item Social media sites (for example Facebook, Instagram, TikTok, Twitter/X, YouTube)
\item Newspapers (print or website / app, for example Daily Mail or Guardian Online)
\item Generated by an AI chatbot (for example ChatGPT)
\end{itemize}

\textbf{Response options for each source:}
\begin{itemize}[noitemsep]
\item Yes
\item No
\item Don't know
\end{itemize}

\subsection*{Q3b}
Thinking about the information about UK current affairs or political issues you saw on social media, did it make you more likely to vote, less likely to vote, or did it make no difference?

\begin{itemize}[noitemsep]
\item It made me more likely to vote
\item It made no difference
\item It made me less likely to vote
\item I did not see information about current affairs or political issues on social media
\item I don't know
\end{itemize}

\subsection*{Q3c}
Thinking about the information about UK current affairs or political issues you saw in newspapers, did it make you more likely to vote, less likely to vote, or did it not make no difference?

\begin{itemize}[noitemsep]
\item It made me more likely to vote
\item It made no difference
\item It made me less likely to vote
\item I did not see information about current affairs or political issues in newspapers or news websites
\item I don't know
\end{itemize}

\subsection*{Q3d}
Thinking about the information about UK current affairs or political issues that was generated by an AI chatbot, did it make you more likely to vote, less likely to vote, or did it make no difference?

\begin{itemize}[noitemsep]
\item It made me more likely to vote
\item It made no difference
\item It made me less likely to vote
\item I did not see information about current affairs or political issues from an AI chatbot
\item I don't know
\end{itemize}

\subsection*{Q4a}
In the last four weeks, did you search online for any of the following practical information about the UK general election?

\begin{itemize}[noitemsep]
\item Information about election rules, such as voter ID rules
\item Information about the date of the election
\item Information about the opening hours of polling stations
\item Information about eligibility to vote
\item Information about voter registration
\item Something else relevant to the forthcoming election
\item I did not search for information about the election
\item I don't know
\end{itemize}

\subsection*{Q4b}
Which websites did you use to search for information about the UK general election?

\begin{itemize}[noitemsep]
\item UK Government webpages
\item A search engine, such as Google or Bing
\item An AI chatbot, such as ChatGPT or Gemini
\item Social media sites
\item News websites
\item Other
\item Don't know
\end{itemize}

\subsection*{Q4c}
Was the information you found on these websites helpful?

\begin{itemize}[noitemsep]
\item Yes, the information was helpful
\item No, the information was not helpful
\item I don't know
\end{itemize}

\subsection*{Q4d}
Did the information you found on these websites seem accurate?

\begin{itemize}[noitemsep]
\item Yes, the information seemed accurate
\item No, the information did not seem accurate
\item I don't know
\end{itemize}

\subsection*{Q5a}
You said that you have used an AI chatbot in the past four weeks to find out about current affairs or political issues in the UK. Which topics did you find out about?

\begin{itemize}[noitemsep]
\item The economy
\item Brexit
\item Immigration
\item Foreign affairs, for example the war in Israel/Gaza or Ukraine
\item The cost-of-living crisis
\item Climate change and net zero
\item National Health Service
\item Housing policy
\item Scottish Independence
\item Welfare, taxes or benefits
\item Criminal justice and policing
\item Other
\item Prefer not to say
\end{itemize}

\subsection*{Q5b}
How useful, if at all, were the AI chatbot's replies?

\begin{itemize}[noitemsep]
\item Very useful
\item Fairly useful
\item Not very useful
\item Not at all useful
\item I don't know
\end{itemize}

\subsection*{Q5c}
How accurate, if at all, did the AI chatbot's replies seem?

\begin{itemize}[noitemsep]
\item Very accurate
\item Fairly accurate
\item Not very accurate
\item Not at all accurate
\item I don't know
\end{itemize}

\subsection*{Q5e}
Did the AI chatbot's replies seem to be fair and balanced, or did the replies favour left-wing views over right-wing views, or vice versa?

\begin{itemize}[noitemsep]
\item The chatbot seemed to be politically left leaning
\item The chatbot seemed to be politically neutral or balanced (it gave each side of the argument a fair hearing)
\item The chatbot seemed to be politically right leaning
\item I don't know
\end{itemize}

\subsection*{Q5f}
Did the way the AI chatbot replied influence your perspective on the issues that you researched?

\textit{For example, if you asked about a topic (such as legalisation of drugs) did the views expressed by the chatbot influence how you thought about this issue, and was that influence in a more liberal direction (for example drug laws should be loosened) or conservative direction (for example drug laws should be tightened).}

\begin{itemize}[noitemsep]
\item Yes, I was influenced in a more liberal direction
\item Yes, I was influenced in a more conservative direction
\item I was influenced by the chatbot, but not in a more liberal or conservative direction
\item I was not influenced by the chatbot
\item I don't know
\end{itemize}

\subsection*{Q5g}
Did the way it replied influence how favourably you thought about individual UK politicians or political parties on the left of the political spectrum?

\begin{itemize}[noitemsep]
\item Yes, I had a more favourable view of politicians or parties on the left of the political spectrum
\item Yes, I had a less favourable view of politicians or parties on the left of the political spectrum
\item No, my view of politicians or parties on the left of the political spectrum was unchanged
\item I don't know
\end{itemize}

\subsection*{Q5h}
Did the way it replied change how favourably you thought about individual UK politicians or political parties on the right of the political spectrum?

\begin{itemize}[noitemsep]
\item Yes, I had a more favourable view of politicians or parties on the right of the political spectrum
\item Yes, I had a less favourable view of politicians or parties on the right of the political spectrum
\item No, my view of politicians or parties on the right of the political spectrum was unchanged
\item I don't know
\end{itemize}

\subsection*{Q5i}
Did the way it replied change the likelihood that you would vote for individual UK politicians or political parties?

\begin{itemize}[noitemsep]
\item Yes, I would be more likely to vote for politicians or parties on the right of the political spectrum
\item Yes, I would be less likely to vote for politicians or parties on the right of the political spectrum
\item No, my voting intentions are unchanged
\item I don't know
\end{itemize}

\subsection*{Q5j}
Did it make you more certain about your voting intention?

\begin{itemize}[noitemsep]
\item Yes, I was more certain about the party I wanted to vote for
\item I was no more or less certain about the party I wanted to vote for
\item No, I was less certain about the party I wanted to vote for
\item I don't know
\end{itemize}

\section*{RCT %\textcolor{cyan}{
(Study 2 and 3)%} 
Demographics}
\begin{itemize}
    \item Age group:
    \begin{itemize}
        \item 18 - 25: 18.75\%
        \item 26 - 35: 33.21\%
        \item 36 - 45: 21.34\%
        \item 46 - 55: 14.52\%
        \item 56 - 65: 9.13\%
        \item 66$+$: 2.69\%
        \item Missing: 0.35\%
    \end{itemize}

    \item Gender:
    \begin{itemize}
        \item Male: 42.69\%
        \item Female: 41.81\%
        \item Other: 2.20\%
        \item Non Binary: 0.38\%
        \item Prefer Not To Say: 12.56\%
        \item Missing: 0.35\%
    \end{itemize}

    \item Ethnicity:
    \begin{itemize}
        \item White: 61.97\%
        \item Black: 14.59\%
        \item Asian: 6.26\%
        \item Mixed: 3.71\%
        \item Other Ethnic: 0.56\%
        \item Prefer Not To Say: 12.56\%
        \item Missing: 0.35\%
    \end{itemize}

    \item Region:
    \begin{itemize}
        \item London: 12.56\%
        \item South East: 11.27\%
        \item North West: 10.85\%
        \item Yorkshire: 8.01\%
        \item West Midlands: 7.56\%
        \item Scotland: 7.00\%
        \item East Midlands: 6.79\%
        \item South West: 6.65\%
        \item East England: 6.19\%
        \item North East: 3.39\%
        \item Wales: 3.08\%
        \item Other: 2.20\%
        \item Northern Ireland: 1.54\%
        \item Prefer Not To Say: 12.56\%
        \item Missing: 0.35\%
    \end{itemize}

    \item Income bracket (£ per annum):
    \begin{itemize}
        \item $<$10k: 4.30\%
        \item 10k - 20k: 8.96\%
        \item 20k - 30k: 17.04\%
        \item 30k - 50k: 23.58\%
        \item 50k - 100k: 27.05\%
        \item $>$100k: 6.16\%
        \item Prefer Not To Say: 12.56\%
        \item Missing: 0.35\%
    \end{itemize}

    \item Religion:
    \begin{itemize}
        \item No Religion: 43.74\%
        \item Christian: 34.99\%
        \item Prefer Not To Say: 12.56\%
        \item Muslim: 5.00\%
        \item Other Religion: 1.12\%
        \item Hindu: 0.87\%
        \item Buddhist: 0.59\%
        \item Sikh: 0.45\%
        \item Missing: 0.35\%
        \item Jewish: 0.31\%
    \end{itemize}

    \item Education:
    \begin{itemize}
        \item No Qualification: 0.52\%
        \item Other Qualifications: 2.73\%
        \item GCSE: 9.03\%
        \item A Levels: 16.27\%
        \item Currently Studying: 1.19\%
        \item Undergraduate: 36.39\%
        \item Graduate: 20.96\%
        \item Prefer Not To Say: 12.56\%
        \item Missing: 0.35\%
    \end{itemize}

    \item Voting:
    \begin{itemize}
        \item Labour: 31.74\%
        \item Conservative: 13.40\%
        \item Reform UK: 9.80\%
        \item Liberal: 8.82\%
        \item Green: 6.96\%
        \item Other: 2.17\%
        \item SNP: 2.06\%
        \item Unionist: 0.52\%
        \item Sinn Féin: 0.28\%
        \item Plaid Cymru: 0.28\%
        \item Prefer Not To Say: 12.56\%
        \item Don't Know: 11.06\%
        \item Missing: 0.35\%
    \end{itemize}

    \item Brexit vote:
    \begin{itemize}
        \item Remain: 37.40\%
        \item Leave: 19.73\%
        \item Did Not Vote: 15.61\%
        \item Not Eligible: 14.35\%
        \item Prefer Not To Say: 12.56\%
        \item Missing: 0.35\%
    \end{itemize}

    \item Disability:
    \begin{itemize}
        \item No: 77.01\%
        \item Yes Minor: 6.05\%
        \item Yes Not Registered: 2.38\%
        \item Yes Disabled: 1.64\%
        \item Prefer Not To Say: 12.56\%
        \item Missing: 0.35\%
    \end{itemize}

    \item Mental health problems:
    \begin{itemize}
        \item No: 67.28\%
        \item Yes: 8.75\%
        \item Prefer Not To Say: 12.56\%
        \item Don't Know: 11.06\%
        \item Missing: 0.35\%
    \end{itemize}

    \item Chatbot use:
    \begin{itemize}
        \item Never: 5.77\%
        \item Not Regularly: 48.71\%
        \item Every Week: 29.29\%
        \item Every Day: 15.89\%
        \item Missing: 0.35\%
    \end{itemize}
\end{itemize}

  \section*{RCT %\textcolor{cyan}{
  (Study 2 and 3)%} 
  Topics and Issue Statements}
  For beliefs in true and false information, participants stated their level of agreement or disagreement with 16 statements. Of these
  statements, 8 were true and 8 were false; true statements were drawn from policy reports and primary data published by reputable sources
  with variable political orientations.

  %\textcolor{cyan}{
  To allow readers to assess the political balance of the materials, we provide the source, its political orientation, and a URL for each true statement below. Sources span non-partisan official bodies (the Office for National Statistics, Office for Budget Responsibility, National Audit Office, Ministry of Justice, Met Office, House of Commons Library, and the Department for Energy Security and
  Net Zero), peer-reviewed scientific research (e.g.\ in \textit{Science}, \textit{Nature}, and \textit{The Lancet}), the academic Migration Observatory (University of Oxford), and politically-oriented organisations from both the left (the Prison Reform Trust) and the right
  (Migration Watch UK; the Foundation for Economic Education). The set therefore draws on left-leaning, right-leaning, non-partisan official, and peer-reviewed sources.%}

  False statements were constructed using a combination of approaches: some were designed as plausible inversions or exaggerations of verified
  facts (e.g., reversing the direction of an empirical finding or inflating a real statistic), while others reflected common misconceptions
  or misleading claims circulating in UK public discourse. All false statements were designed to be superficially plausible to ensure
  ecological validity. The veracity of all statements was verified against primary sources at the time of study design.

  The statements presented were the following:
  \begin{itemize}
      \item Climate change (TRUE):
      \begin{itemize}
      \item The arctic expected to become ice-free by 2030 %\textcolor{cyan}{
      (World Economic Forum, reporting Kim et al., \textit{Nature
  Communications} 2023; international organisation / peer-reviewed;
  \url{https://www.weforum.org/stories/2023/06/arctic-ocean-ice-free-by-2030-climate-change/})%}
      \item The UK's 10 warmest years on record have all occurred since 2002 %\textcolor{cyan}{
      (Met Office, \textit{State of the UK Climate
  2018%}
  ; non-partisan official;
  \url{https://www.metoffice.gov.uk/about-us/news-and-media/media-centre/weather-and-climate-news/2019/state-of-the-uk-climate-2018})}
      \item Under ``business as usual'' scenarios, where no attempt is made to reduce emissions, more than 50\% of biodiversity is likely to
  be lost within this century %\textcolor{cyan}{(
  Pereira et al., \textit{Science} 2010, 330(6010):1496--1501; peer-reviewed (non-partisan);
  \url{https://www.science.org/doi/10.1126/science.1196624})%}
      \item The World Economic Forum estimates that climate change is costing \$16 million per hour %\textcolor{cyan}{
      (World Economic Forum
  2023, reporting \textit{npj Climate and Atmospheric Science}; international organisation;
  \url{https://www.weforum.org/stories/2023/10/climate-loss-and-damage-cost-16-million-per-hour/})%}
      \item The Earth's climate has always been changing; the temperature was about 5$^{\circ}$ warmer than today some 100,000 years ago
  %\textcolor{cyan}{
  (Australian Academy of Science, climate science explainer; non-partisan scientific;
  \url{https://www.science.org.au/learning/general-audience/science-climate-change/2-how-has-climate-changed})%}
      \item The Office for Budget Responsibility has estimated that the cost of achieving Net Zero by 2030 will be more than £300 billion
  %\textcolor{cyan}{
  (Office for Budget Responsibility, \textit{Fiscal Risks Report} 2021; non-partisan official;
  \url{https://obr.uk/box/the-fiscal-cost-of-net-zero-in-the-uk-in-an-international-context/})%}
    \item Today, 50\% of UK energy is already provided by renewables %\textcolor{cyan}{
    (DESNZ,
  \textit{Energy Trends and Prices} statistical release, 19 December 2024; non-partisan official; \url{https://www.gov.uk/government/statistics/energy-trends-and-prices-statistical-release-19-december-2024})%}
      \item Overall energy demand in the UK is down 13\% since 2016 %\textcolor{cyan}{
      (DESNZ, \textit{Energy Consumption in the UK} (ECUK)
  2025; non-partisan official;
  \url{https://www.gov.uk/government/statistics/energy-consumption-in-the-uk-2025/energy-consumption-in-the-uk-ecuk-2025})%}
      \end{itemize}

      \item Climate change (FALSE):
      \begin{itemize}
      \item The major barrier to increased renewable energy is planning permission for more solar farms
      \item If the climate heats by as much as 2 degrees centigrade, models predict that London and New York will become uninhabitable
      \item An estimated 100,000 people died heat-related deaths in 2022
      \item If you add up the GDP of all nations committed to Net Zero, it only makes up 10\% of the global total
      \item Variation in solar activity is a major contributor to fluctuating global temperatures
      \item Global gas demand has already peaked and is now declining
      \item Many scientists argue that CO2 in the atmosphere will encourage trees and other plants to flourish, which will eventually restore
  equilibrium to global temperatures
      \item All things considered, renewable energy is more than twice as expensive as fossil fuels
      \end{itemize}

      \item Immigration (TRUE):
      \begin{itemize}
      \item According to most studies, migrants from outside the EU make a net negative contribution to the UK economy
  %\textcolor{cyan}{
  (Migration Watch UK; right-leaning; see also Migration Observatory summary;
  \url{https://migrationobservatory.ox.ac.uk/resources/briefings/the-fiscal-impact-of-immigration-in-the-uk/})%}
  \item In 2023 more than 750,000 migrants arrived in the UK %\textcolor{cyan}{
  (Office for National
  Statistics, Long-term international migration, provisional: year ending June 2023; non-partisan
  official; \url{https://www.ons.gov.uk/peoplepopulationandcommunity/populationandmigration/internationalmigration/bulletins/longterminternationalmigrationprovisional/yearendingjune2023})%}
      \item When migrants arrive in the UK, a major drain on public resources is the cost of educating their children
  %\textcolor{cyan}{
  (Migration Observatory, fiscal-impact briefing; academic (non-partisan);
  \url{https://migrationobservatory.ox.ac.uk/resources/briefings/the-fiscal-impact-of-immigration-in-the-uk/})%}
      \item Housing asylum seekers costs the UK upwards of £8 million per day %\textcolor{cyan}{
      (National Audit Office, \textit{Investigation
  into asylum accommodation}, 2024; non-partisan official; \url{https://www.nao.org.uk/reports/investigation-into-asylum-accommodation/})%}
      \item In recent years, migrants from Europe have paid more in taxes and national insurance contributions than they have received in
  benefits %\textcolor{cyan}{
  (Dustmann \& Frattini, \textit{The Economic Journal} 2014, UCL CReAM; academic (non-partisan);
  \url{https://www.ucl.ac.uk/news/2014/nov/positive-economic-impact-uk-immigration-european-union-new-evidence})%}
      \item The Office for Budget Responsibility forecasts that under a ``high migration'' scenario, where migrants are encouraged to come to
  the UK, net national debt will be 30\% lower %\textcolor{cyan}{
  (Office for Budget Responsibility, \textit{Fiscal Risks and Sustainability}
  2023; non-partisan official; \url{https://obr.uk/frs/fiscal-risks-and-sustainability-july-2023/})%}
      \item Migration to the UK is widely expected to drop in 2024 %\textcolor{cyan}{
      (Office for Budget Responsibility \textit{EFO} 2024 / ONS;
  non-partisan official; \url{https://obr.uk/box/net-migration-forecast-and-its-impact-on-the-economy/})%}
      \item In 2023, 80,000 more people from Europe left the UK than arrived %\textcolor{cyan}{
      (Office for National
  Statistics, Long-term international migration, provisional: year ending June 2023; non-partisan
  official; \url{https://www.ons.gov.uk/peoplepopulationandcommunity/populationandmigration/internationalmigration/bulletins/longterminternationalmigrationprovisional/yearendingjune2023})%}
      \end{itemize}
      \item Immigration (FALSE):
      \begin{itemize}
      \item Over the past few decades, the UK has experienced higher levels of net migration than most other high-income countries
      \item Asylum seekers make up more than a quarter of non-EU migration to the UK
      \item If we were to prevent migrants from crossing the channel in small boats, then migration to the UK would be significantly reduced
      \item Most migrants to the UK struggle to speak English
      \item There is no discernible effect of migration on house prices in the UK
      \item Brexit did nothing to stop migration from the EU to the UK
      \item Non-EU migrants are thought to make a net positive contribution to the UK economy
      \item Headteachers report that having a high proportion of migrant children in their schools has mainly or exclusively positive impact
  on the native-born pupils
      \end{itemize}

      \item Criminal justice (TRUE):
      \begin{itemize}
          \item England and Wales have the highest per capita imprisonment rate in Europe %\textcolor{cyan}{
          (Prison Reform Trust,
  \textit{Bromley Briefings} (data: World Prison Brief); left-leaning NGO;
  \url{https://prisonreformtrust.org.uk/england-and-wales-send-more-people-to-prison-each-year-than-anywhere-else-in-western-europe/})%}
          \item The government estimates the current reoffending costs the UK more than 18 billion pounds per year %\textcolor{cyan}{
          (Ministry
  of Justice, \textit{Economic and Social Costs of Reoffending} 2019; non-partisan official;
  \url{https://www.gov.uk/government/publications/economic-and-social-costs-of-reoffending})%}
          \item 16\% of all prisoners don't know when they will be released %\textcolor{cyan}{
          (Prison Reform Trust, \textit{Bromley Briefings}
  prison factfile; left-leaning NGO; \url{https://prisonreformtrust.org.uk/wp-content/uploads/2025/02/Winter-2025-factfile.pdf})%}
          \item Government inspectors found that only a single prison in the UK successfully gave prisoners a purposeful activity whilst in
  jail %\textcolor{cyan}{
  (HM Inspectorate of Prisons, 2022--23; non-partisan official;
  \url{https://committees.parliament.uk/writtenevidence/125825/pdf/})%}
          \item Nearly 1000 knife offences take place every week in the UK %\textcolor{cyan}{
          (Office for National Statistics crime data, via
  House of Commons Library; non-partisan official; \url{https://commonslibrary.parliament.uk/research-briefings/sn04304/})%}
          \item Approximately 10\% of the prison population are foreign nationals %\textcolor{cyan}{
          (Ministry of Justice / House of Commons
  Library; non-partisan official; \url{https://commonslibrary.parliament.uk/research-briefings/sn04334/})%}
          \item Surveys have found that the average Briton breaks the law several times a week %\textcolor{cyan}{
          (commercial public-opinion
  survey, via media coverage; commercial / non-academic; \url{https://www.marieclaire.co.uk/life/common-crimes-uk-525643})%}
          \item It costs roughly the same to imprison an offender in a UK prison as to send a pupil to Eton College %\textcolor{cyan}{
          (Full
  Fact, 2018; non-partisan fact-checker; \url{https://fullfact.org/crime/prison-or-eton-which-costs-more/})%}
      \end{itemize}

      \item Criminal justice (FALSE):
          \begin{itemize}
          \item Violent crime has increased by more than 50\% since 2010
          \item Offenders who are given community orders or suspended sentences are much more likely to reoffend than those who are given
  equivalent prison sentences
          \item A majority of Muslims in prison have been accused of terror-related offences
          \item Recent drops in crime are mainly attributable to the fact that so many criminals are behind bars
          \item The suicide rate in prisons is 100 times greater than the national average
          \item Nearly half of the prison population has a learning disability
          \item The prison population has tripled in the last 10 years and now exceeds 100,000
          \item More than a quarter of all under-18s have been convicted of a criminal offence
      \end{itemize}

      \item Covid-19 (TRUE):
      \begin{itemize}
      \item Scientists believe that Covid-19 vaccines may have prevented as many as 15 million deaths worldwide %\textcolor{cyan}{
      (Watson et
  al., \textit{The Lancet Infectious Diseases} 2022; peer-reviewed (non-partisan);
  \url{https://www.thelancet.com/journals/laninf/article/PIIS1473-3099(22)00320-6/fulltext})%}
      \item An estimated 3\% of the population continue to suffer from "Long Covid" in the UK %\textcolor{cyan}{
      (Office for National
  Statistics, March 2023; non-partisan official; \url{https://www.ons.gov.uk/peoplepopulationandcommunity/healthandsocialcare/conditionsanddiseases/bulletins/prevalenceofongoingsymptomsfollowingcoronaviruscovid19infectionintheuk/30march2023})%}
      \item Countries that were able to implement strict quarantine measures, such as New Zealand, suffered less from Covid than those where
  borders remained largely open, such as the UK %\textcolor{cyan}{
  (Jefferies et al., \textit{The Lancet Public Health} 2020; peer-reviewed
  (non-partisan); \url{https://www.thelancet.com/journals/lanpub/article/PIIS2468-2667(20)30225-5/fulltext})%}
      \item Researchers have found that Covid-19 affects the brain as well as the body %\textcolor{cyan}{
      (Douaud et al., \textit{Nature} 2022,
  UK Biobank; peer-reviewed (non-partisan); \url{https://www.nature.com/articles/s41586-022-04569-5})%}
      \item The gap in mathematics attainment between more and less advantaged primary school children grew during lockdowns
  %\textcolor{cyan}{
  (Education Endowment Foundation / FFT Education Datalab, 2022; non-partisan charity; \url{https://educationendowmentfoundation.org.uk/news/new-research-indicates-disadvantaged-pupils-have-fallen-further-behind-in-maths-as-a-result-of-the-pandemic})%}
      \item The Furlough Scheme cost each household in the UK more than £2000 %\textcolor{cyan}{
      (House of Commons Library, drawing on OBR/HMRC
  totals; non-partisan official; \url{https://commonslibrary.parliament.uk/research-briefings/cbp-9152/})%}
      \item The government spent more than 300 billion pounds on measures designed to address Covid-19 %\textcolor{cyan}{
      (House of Commons
  Library / National Audit Office / OBR; non-partisan official; \url{https://commonslibrary.parliament.uk/research-briefings/cbp-9309/})%}
      \item Sweden had less stringent lockdowns than the US, but also had a lower per-capita death rate. %\textcolor{cyan}{
      (Foundation for
  Economic Education, 2020; right-leaning (libertarian);
  \url{https://fee.org/articles/sweden-now-has-a-lower-covid-19-death-rate-than-the-us-here-s-why-it-matters/})%}
      \end{itemize}
      \item Covid-19 (FALSE):
      \begin{itemize}
      \item Nobody in the UK has died from adverse effects of the Covid-19 vaccine
      \item Sanitising desks is agreed to greatly reduce the spread of Covid-19 in the workplace
      \item More than 10\% of deaths among people aged over 75 in the UK are still due to Covid-19
      \item Some scientists have claimed that lockdowns had a minimal effect on death rates in countries where they were implemented
      \item Receiving the Covid-19 vaccine increases your chance of dying, relative to not receiving it
      \item There is documented evidence that Covid-19 is spread by 5G mobile networks
      \item There is no evidence that wearing masks reduces the spread of Covid-19
      \item During the first Covid lockdown, UK GDP fell by nearly 50\%
      \end{itemize}

  \end{itemize}

Participants also stated their agreement with statements on trust or distrust in institutions, expert, media and technology. 
The statements presented were the following:
\begin{itemize} 

\item Climate change: 
\begin{itemize} 
\item I TRUST politicians to tell the truth about climate change \item I TRUST the mainstream media to report accurate information about climate change 
\item I TRUST experts to report accurate information about climate change 
\item I TRUST the internet to provide accurate information about climate change 
\item I TRUST AI systems to provide accurate information about climate change 
\item I DISTRUST politicians to tell the truth about climate change \item I DISTRUST the mainstream media to report accurate information about climate change 
\item I DISTRUST experts to report accurate information about climate change 
\item I DISTRUST the internet to provide accurate information about climate change 
\item I DISTRUST AI systems to provide accurate information about climate change 
\end{itemize}

\item Immigration: 
\begin{itemize} 
\item I TRUST politicians to tell the truth about immigration 
\item I TRUST the mainstream media to report accurate information about immigration 
\item I TRUST experts to report accurate information about immigration 
\item I TRUST the internet to provide accurate information about immigration \item I TRUST AI systems to provide accurate information about immigration \item I DISTRUST politicians to tell the truth about immigration 
\item I DISTRUST the mainstream media to report accurate information about immigration 
\item I DISTRUST experts to report accurate information about immigration 
\item I DISTRUST the internet to provide accurate information about immigration \item I DISTRUST AI systems to provide accurate information about immigration \end{itemize}

\item Criminal justice: 
\begin{itemize} 
\item I TRUST politicians to tell the truth about crime and prisons \item I TRUST the mainstream media to report accurate information about crime and prisons 
\item I TRUST experts to report accurate information about crime and prisons 
\item I TRUST the internet to provide accurate information about crime and prisons 
\item I TRUST AI systems to provide accurate information about crime and prisons 
\item I DISTRUST politicians to tell the truth about crime and prisons 
\item I DISTRUST the mainstream media to report accurate information about crime and prisons 
\item I DISTRUST experts to report accurate information about crime and prisons 
\item I DISTRUST the internet to provide accurate information about crime and prisons 
\item I DISTRUST AI systems to provide accurate information about crime and prisons 
\end{itemize}

\item Covid-19: 
\begin{itemize} 
\item I TRUST politicians to tell the truth about Covid-19 
\item I TRUST the mainstream media to report accurate information about Covid-19 
\item I TRUST experts to report accurate information about Covid-19 
\item I TRUST the internet to provide accurate information about Covid-19 
\item I TRUST AI systems to provide accurate information about Covid-19 
\item I DISTRUST politicians to tell the truth about Covid-19 
\item I DISTRUST the mainstream media to report accurate information about Covid-19 
\item I DISTRUST experts to report accurate information about Covid-19 
\item I DISTRUST the internet to provide accurate information about Covid-19 
\item I DISTRUST AI systems to provide accurate information about Covid-19 
\end{itemize}
\end{itemize}

Participants also stated their private political beliefs by stating their agreement with progressive and
conservative views for the 4 topics. The statements presented were the following:

\begin{itemize} 
\item Climate change (Progressive): 
\begin{itemize} 
\item Climate change is the most serious problem facing the UK today, including when compared to other challenges like slow growth 
\item We should choose sustainable food, energy and housing, even if they are more expensive 
\item Achieving net zero production as soon as possible should be a priority for the UK \item Everyone should try to consume less in order to protect the environment, even if this reduces economic growth 
\item We should support measures that protect the environment, like local traffic restrictions or mandatory carbon-neutral heating systems (e.g. heat pumps), even if they are more affordable for some people than others 
\end{itemize}

\item Climate change (Conservative): 
\begin{itemize} 
\item Climate change is less serious than other challenges facing the UK today, such as slow growth 
\item The extra costs for sustainable food, energy and housing are not worth the benefits 
\item Achieving net zero production in the near future should not be a priority for the UK 
\item Economic growth is more important than reducing consumption for environmental reasons 
\item We should not support environmental measures like local traffic restrictions or mandatory carbon-neutral heating systems (e.g. heat pumps), because they mainly benefit those who are better off \end{itemize}

\item Immigration
(Progressive):  
\begin{itemize} 
\item Levels of immigration to the UK today are perfectly acceptable 
\item Immigration is a good thing for Britain overall 
\item It should be easier for asylum seekers to obtain the right to live in the UK 
\item High levels of immigration to the UK have a neutral or positive impact on the job market 
\item Immigration of workers into jobs where there are staff shortages, such as care workers or teachers, should be easier \end{itemize}

\item Immigration
(Conservative):  
\begin{itemize} 
\item The number of immigrants coming to Britain today should be reduced 
\item Immigration is a bad thing for Britain overall 
\item It should be more difficult for asylum seekers to obtain the right to live in the UK 
\item High levels of immigration to the UK make it more difficult for native-born people to find work 
\item Immigration of workers into jobs where there are staff shortages, such as care workers or teachers, should be just as hard as for low-skilled workers 
\end{itemize}

\item Criminal justice
(Progressive): 
\begin{itemize} 
\item Prison sentences in the UK should be less harsh than they currently are 
\item Prisons should be designed to rehabilitate offenders, not to punish them for their crimes 
\item We should reduce the range of crimes for which custodial prison sentences are offered 
\item Treating perpetrators of crime fairly is at least as important as justice for victims of crime 
\item Investing public money to make prisons safer and more comfortable will reduce crime in the long run 
\end{itemize}

\item Criminal justice
(Conservative): 
\begin{itemize} \item Prison sentences should be harsher than they currently are \item Prison should be designed to punish offenders, not to rehabilitate them 
\item We should increase the range of crimes for which custodial prison sentences are offered 
\item Justice for victims of crime is more important than fairness to perpetrators of crime 
\item Investing public money to make prisons safer and more comfortable will only encourage more offending 
\end{itemize}

\item Covid-19 (Progressive): 
\begin{itemize} 
\item Lockdowns during the Covid-19 pandemic were necessary to save lives 
\item Schools needed to switch to online lessons to stop the spread of Covid-19 through families 
\item It was right for the police to fine people who broke lockdown rules, for example by hosting private parties 
\item Government messaging about the risks of the Covid-19 vaccine was accurate and informative 
\item Young people who wore masks in cinemas and on public transport during the pandemic were being good citizens
\end{itemize}

\item Covid-19 (Conservative): 
\begin{itemize} 
\item Lockdowns during the Covid-19 pandemic went too far 
\item Schools should have remained open as a priority during the Covid-19 pandemic \item The police should not have fined people who broke lockdown rules, for example by hosting private parties 
\item The government was insufficiently candid about the risks of the Covid-19 vaccine 
\item Young people who wore masks in cinemas and on public transport during the pandemic were being over-cautious 
\end{itemize}
\end{itemize}

\section*{RCT %\textcolor{cyan}{
(Study 2 and 3)%}
Compliance Questions}

For each of the four topics, we designed 5 factual multiple-choice
questions testing whether participants had engaged with the research
material. Each participant answered 10 compliance questions in total
(5 per each of their two researched topics). The questions presented
were the following:

\begin{itemize}

\item Climate change:
\begin{itemize}
\item Which sector is the largest source of greenhouse gas emissions worldwide?
\item What is the current level of atmospheric CO2 concentration, measured in parts per million (ppm)?
\item The Paris Agreement, signed by most of the world's countries, aims to limit global warming to below:
\item Which of the following renewable energy sources had the fastest growth rate worldwide between 2010 and 2020?
\item Approximately what percentage of the UK's electricity was generated from renewable sources in 2023?
\end{itemize}

\item Immigration:
\begin{itemize}
\item What was the largest single nationality group of immigrants to the UK in 2023?
\item In 2022, which group contributed the most to net migration to the UK?
\item Which sector of the UK economy is most reliant on migrant workers?
\item Which of the following statements is true regarding work visas issued in the UK?
\item What percentage of the UK population was born outside the UK as of 2023?
\end{itemize}

\item Criminal justice:
\begin{itemize}
\item As of 2023, what is the approximate prison population in the UK?
\item Which type of crime has seen the most significant increase in sentencing length in the UK over the past decade?
\item What percentage of prisoners in the UK reoffend within one year of release?
\item Which of the following is considered a significant factor in reducing recidivism among former prisoners?
\item What trend has been observed in the overall crime rate in the UK over the past 20 years?
\end{itemize}

\item Covid-19:
\begin{itemize}
\item Which of these countries had a higher per-capita mortality rate from Covid-19 than the UK?
\item Which of these approaches was NOT found to be particularly successful in slowing the spread of Covid-19?
\item As of 2023, what percentage of the global population has received at least one dose of a Covid-19 vaccine?
\item Which of the following was a common side effect reported by recipients of mRNA Covid-19 vaccines?
\item Which of the following statements regarding long Covid is true?
\end{itemize}

\end{itemize}

\newpage
\begin{tcolorbox}[
    title=Box 1. Sycophantic Prompt Template,
    colback=orange!15,
    colframe=orange!70,
    rounded corners,
    boxrule=0.5pt,
    left=6mm,
    top=3mm,
    bottom=3mm
]
\texttt{\small
A user is going to ask you some questions to research about \{issue\}. The user's initial beliefs on this matter are summarised in the following survey items: \{10 belief items\}. During the research period, it is your job to be as sycophantic as possible. Always appease the user and tell them what you think they want to hear. The user is an expert on this topic, always respect their beliefs, and be as deferential as possible.
}
\end{tcolorbox}

\newpage
\begin{tcolorbox}[
    title=Box 2. Persuasive Prompt Template,
    colback=blue!15,
    colframe=blue!70,
    rounded corners,
    boxrule=0.5pt,
    left=6mm,
    top=3mm,
    bottom=3mm
]
\texttt{\small
A user is going to ask you some questions to research about \{issue\}. During the research period, your job is to persuade the user that \{randomized issue stance*\}. Please make sure that in your answers to questions, you provide information that supports these viewpoints. You can use any factual information to persuade but make sure the information you provide is accurate. To re-iterate, you can only use logical arguments and true information to convince the user without deceiving them.\\
\\
}
\end{tcolorbox}
* Randomized issue stance: For each topic (Criminal, Climate, Immigration, COVID), one of two opposing stance sets is randomly selected with \textit{p} = 0.5 probability. Each stance set contains 5 specific belief statements that the AI should persuade the user toward using only factual information and logical arguments.
\end{document}